\documentclass[useAMS, usenatbib]{mn2e}

\usepackage{amssymb}
\usepackage{graphicx}
\usepackage{threeparttable}
\usepackage{multirow}
\usepackage{longtable}
\usepackage{subfig}
\usepackage{url}
\usepackage{epsfig}

\newcommand{\mum}{\ifmmode{\rm \mu m}\else{$\mu$m }\fi}

\newcommand{\chisq}{\ifmmode{\chi^{2} }\else{$\chi^2$}\fi}
\newcommand{\rchisq}{\ifmmode{\chi^{2} }\else{$\chi^2_\nu$}\fi}

\title[Alumina abundance of AGB stars in the LMC]
{Modelling the alumina abundance of oxygen-rich evolved stars in the Large Magellanic Cloud}

\author[O. C. Jones et al.]   
{O.~C.~Jones,$^1$ \thanks{E-mail:ojones@jb.man.ac.uk}
  F.~Kemper,$^{2}$ 
  S.~Srinivasan,$^{2}$
  I.~McDonald,$^1$
  G.~C.~Sloan$^{3}$
and
\newauthor
 A.~A.~Zijlstra$^1$ \\
$^1$ Jodrell Bank Centre for Astrophysics, Alan Turing Building, School of Physics \& Astronomy, The University of Manchester, \\
Oxford Road, Manchester, M13 9PL, UK.\\
$^{2}$ Institute of Astronomy and Astrophysics, Academia Sinica, 11F ASMAB, NTU, No.~1, Sec.~4, Roosevelt Rd, Taipei 10617, Taiwan, R.O.C.\\
$^{3}$ Department of Astronomy, Cornell University, Ithaca, NY 14853, USA.}

\begin{document}

\date{Accepted XXXX XXXXXX XX. Received XXXX XXXXXX XX}

\pagerange{\pageref{firstpage}--\pageref{lastpage}} \pubyear{2014}

\maketitle

\label{firstpage}

\begin{abstract}

In order to determine the composition of the dust in the circumstellar envelopes of oxygen-rich asymptotic giant branch (AGB) stars we have computed a grid of {\sc modust} radiative-transfer models for a range of dust compositions, mass-loss rates, dust shell inner radii and stellar parameters. We compare the resulting colours with the observed oxygen-rich AGB stars from the SAGE-Spec Large Magellanic Cloud (LMC) sample, finding good overall agreement for stars with a mid-infrared excess.
We use these models to fit a sample of 37 O-rich AGB stars in the LMC with optically thin circumstellar envelopes, for which 5--35-\mum {\em Spitzer} infrared spectrograph (IRS) spectra and broadband photometry from the optical to the mid-infrared are available. From the modelling, we find mass-loss rates in the range $\sim 8\times10^{-8}$ to $5\times10^{-6}$ M$_{\odot}\ \mathrm{yr}^{-1}$, and we show that a grain mixture consisting primarily of amorphous silicates, with contributions from amorphous alumina and metallic iron provides a good fit to the observed spectra. 
Furthermore, we show from dust models that the {\em AKARI} [11]--[15] versus [3.2]--[7] colour-colour diagram, is able to determine the fractional abundance of alumina in O-rich AGB stars.

\end{abstract}

\begin{keywords}
radiative transfer -- stars: AGB, post-AGB -- circumstellar matter -- dust, extinction -- Magellanic Clouds -- infrared: stars
\end{keywords}

\section{Introduction}

Stars on the asymptotic giant branch (AGB) lose a significant fraction of their matter through slow dense winds at rates of $10^{-10}$ to $10^{-4}$ M$_\odot$ yr$^{-1}$ \citep{Bowen1991}. During this phase of evolution the effective stellar temperature is low enough for molecules to form and dust grains to condense in the circumstellar outflow. 
In general, dust formation occurs in stages known as the `dust condensation sequence', and depends on the physical conditions in the envelope \citep[][]{Tielens1990, GailSed1999, Gail2010}. The mineralogy of the dust is dominated by the chemistry of the envelope (which is set by a star's photospheric C/O atomic ratio and initial elemental abundance), and by the envelope's temperature and density (which determines which condensation products are stable). 

Although not understood in detail, the thermodynamic condensation sequence for oxygen-rich outflows (C/O $<$ 1) predicts that refractory oxides such as alumina (Al$_{2}$O$_{3}$) would be the first astrophysically significant species to form. These grains can exist relatively close to the star ($T_{\rm cond} \sim 1400$ K at pressures relevant to stellar outflows) and may act as seed nuclei for other grains \citep[e.g.][]{Onaka1989, Stencel1990, Sogawa1999}. At larger radii, where the temperature is lower ($T \sim 1000$ K) silicate grains form. 

It is well established that the dust in oxygen-rich outflows consists mainly of silicates  \citep{WoolfNey1969, Hackwell1972, TreffersCohen1974}. However, alumina dust has been detected in the spectra of some low mass-loss rate ($\dot M$) O-rich AGB stars in the Milky Way \citep{Onaka1989, Speck2000, Dijkstra2005}. These stars show a broad low-contrast emission feature in the 7 -- 14 $\mu$m region which is best fit with a blend of amorphous alumina and amorphous silicates.
As the star evolves along the AGB the shape and peak position of this feature changes; silicate dust becomes more important and dominates the 10-$\mu$m emission \citep{Little-Marenin1990,Sloan1995}. At high mass-loss rates the 10-$\mu$m feature is only due to silicates. This sequence has been quantified by \cite{Sloan1995, Sloan1998} using flux ratios in the 9--12 $\mu$m region.

The \emph{Spitzer} Space Telescope has taken spectra of a large number of individual evolved stars in the Large and Small Magellanic Clouds, covering a wide range of colours and magnitudes \citep{Kemper2010}. 
Although the mineralogy of the oxygen-rich AGB stars in the Magellanic Clouds is dominated by amorphous silicates both at low and high mass loss rates \citep{Sargent2010, Riebel2012} some sources also contain a small fraction of crystalline silicates \citep{Jones2012}. There is an apparent absence of Al$_{2}$O$_{3}$ in the \emph{Spitzer} spectra of O-AGB stars in the Magellanic Clouds \citep{Sloan2008}, with the 10-$\mu$m feature of low density (low $\dot M$) winds better reproduced by amorphous silicate grains than the expected mixture of Al$_{2}$O$_{3}$ and silicates. This implies that either alumina dust is depleted in the LMC or that observational biases hinder its detection. 
However, an extensive systematic study into the Al$_{2}$O$_{3}$ content of evolved stars in the LMC has not yet been carried out.

In this study we evaluate the alumina content of oxygen-rich AGB stars in the Magellanic Clouds, by creating a grid of radiative transfer models that explores a range of physical parameters relevant to evolved stars. 
In Section~\ref{model}, we describe how the radiative transfer models for the circumstellar dust shell were computed. Colour-colour diagrams of the grid of models are presented in Section~\ref{results} and will be used to study the alumina dust in a sample of O-rich AGB stars in the Magellanic Clouds in Section~\ref{AluminaLMC}. Finally, we discuss our results in Section~\ref{Discussion} and summarise this study in Section~\ref{conclusions}.

\section{Modelling the circumstellar dust shell}\label{model} 

To determine the relative contributions of alumina dust from the spectra of O-rich AGB stars in the LMC, radiative transfer modelling is required to calculate detailed spectra of circumstellar dust shells. In this paper we use the one-dimensional radiative transfer code \textsc{modust} \citep{Bouwman2000, Bouwman2001, Kemper2001} to evaluate the emergent spectrum from a central star surrounded by a spherically-symmetric dust shell. \textsc{modust} solves the transfer equations for the basic free parameters such as stellar temperature, chemical composition, grain size distribution, etc.~under the constraint of radiative equilibrium, using a Feautrier-type method \citep{Feautrier1964}. It also allows the user to implement multiple dust shells, each consisting of different dust components with an independent grain-size distribution function.  In the following sub-sections we will describe the selection of the input parameters adopted in the modelling.

\subsection{Model input}

\subsubsection{Photospheres}

The models have been calculated using three different synthetic stellar photospheres for O-rich stars of solar-metallicity taken from \cite{Fluks1994}. The spectral types of the central star used in the model grid are: M1 with an effective stellar temperature ($T_{{\rm eff}}$) of 3715 K, M5  ($T_{{\rm eff}}$ = 3396 K) and M9  ($T_{{\rm eff}}$ = 2667 K). These were chosen to reproduce the range of values expected for O-AGB stars.
As luminosity only acts as a scaling factor for the emission and does not affect the spectral shape \citep{Ivezic1997}, a typical luminosity for AGB stars of 7000 $L_{\odot }$ was assumed, resulting in stellar radii ($R_{\star}$) for the central star of 202.3, 242.1 and 392.5 $R_{\odot}$ (respectively).


\begin{table}
 \caption{Model Parameters}
 \label{tab:modelInputAGB}
  \begin{center}
    \begin{tabular}{@{}ll}
      \hline
      \hline
      Parameter                          & Range of Values          \\
      \hline
      \bf{Star}                          &                          \\
       Spectral type                     &    M1, M5, M9            \\
       $L_{{\rm \star}} \,  (L_{\odot})$     &    7000                 \\
       $T_{{\rm eff}}$ (K)                 &    3715,  3396, 2667     \\
       $R_{{\rm \star}} \,  (R_{\odot})$     &   202.3, 242.1, 392.5    \\
                                         &                            \\  
      \bf{Dust Shell Properties}         &                            \\
       $R_{{\rm in}} \,  (R_{{\rm \star}})$   &   2.5, 3, 5, 7.5 and 15  \\
       $R_{{\rm out}}/R_{{\rm in}}$           &     200                  \\
       $v_{{\rm exp}} \, (\rm {kms}^{-1})$  &     10                     \\
       Dust-to-gas ratio ($\Psi$)        &     1/200                   \\                  
       Density profile                   & $\rho(r) \sim r^{-2}$        \\
       $\dot{M}$ (M$_{\odot}\, \mathrm{yr}^{-1}$)  &   $10^{-10}$--$10^{-5}$     \\  
                                        &                               \\  
     \bf{Dust Grain Properties}         &                               \\
       Size distribution                &  MRN                          \\
       Grain shape                      &  CDE                          \\
       $a_{\rm {min}} \, (\mu$m)          &  0.01                          \\ 
       $a_{\rm {max}} \, (\mu$m)          &  1.00                          \\ 
       $q$                              &  3.5                           \\     
                                        &                                \\  
      \bf{Dust Species}                 &   {\bf Reference}               \\
      Amorphous Mg$_2$SiO$_4 $          & \cite{Dorschner1995}            \\
       Am. Al$_2$O$_3$                  &  \cite{Begemann1997};           \\
                                        & \cite{Koike1995}                \\
       Fe                               & \cite{Ordal1988}                \\                 
     \hline
    \hline
    \end{tabular}
  \end{center}
\end{table}

\subsubsection{Dust shell geometry}

The dust is assumed to be distributed in a spherical shell with an $r^{-2}$ density distribution, giving rise to a spherically-symmetric, time-independent stellar wind with a constant outflow velocity. 
The lack of silicate absorption features in the spectra of O-rich AGB stars in the LMC sample and the relative weakness of the mid-IR flux compared to the flux at 1 $\mu$m \citep{Sargent2010} indicates that these stars typically have optically-thin dust shells in the mid-IR, thus limiting the column density along the line of sight. In optically-thin dust shells the geometry of the circumstellar shell has little influence on the appearance of the spectral features, as all grains receive approximately the same amount of (near-IR) stellar photons, and absorption of scattered or emitted radiation is minimal due to the decreasing extinction efficiency of the oxygen-rich dust at longer wavelengths. However, the density profile of the dust shell determines the dust mass distribution over temperature, thus affecting the continuum emission levels and feature strengths. For optically-thick dust shells the geometry of the circumstellar dust would significantly influence the shape of the spectral features and would require a more complex treatment of the density distribution. 

The inner radius of the dust shell ($R_{{\rm in}}$) is determined assuming that the dust temperature at this radius is equal to the condensation temperature of the dust species, and that 

\begin{equation}\label{eq:inner_radius}
      T_{{\rm d}}(r)=T_{\star}\left(\frac{R_{\star}}{2r}\right)^{2/(4+s)},
\end{equation}
with $s \approx 1$ \citep{Olofsson2004}. For amorphous silicates the condensation temperature is  $T_{{\rm cond}} \sim 1000$ K \citep{GailSed1999}, however alumina is expected to condense at higher temperatures  $T_{{\rm cond}} \sim 1400$ K. To be consistent with other modelling efforts \citep[e.g.][]{Heras2005, Groenewegen2006, Sargent2011, Srinivasan2011} we do not include multiple dust shells with separate dust components, each with its own temperature distribution. Instead, to account for this range in condensation temperature, and for dynamical effects such as pulsations,  $R_{{\rm in}}$ is varied for a best fit. We calculate models for $R_{{\rm in}}$ = 2.5, 3, 5, 7.5 and 15 $R_{\star}$. These are similar to the range of $R_{{\rm in}}$ values explored by the O-rich {\sc grams} models \citep{Sargent2011},  based on estimates for the condensation radius by \cite{Hoefner2007}. 

In some instances a given set of model parameters may result in dust temperatures at the inner regions of the shell that are in excess of the condensation temperature. We deem these models to be unphysical and eliminate them from our grid.

The outer radius ($R_{{\rm out}}$) determines the duration of the mass loss and hence the total shell mass, and increases the amount of cold dust. In order to accurately determine the thickness of the shell, far-IR data is required.  The dust temperatures at the outermost regions of the wind only contribute slightly to the mid-IR flux and hence cannot be constrained by our models. For instance, for the M9III model with  $R_{{\rm in}}$  = 7.5 and a total mass-loss rate of $1\times10^{-6}$ M$_{\odot }$ yr$^{-1}$, the 24-\mum flux was only  $\sim$ 4.8 per cent  higher for $R_{{\rm out}} = 1000 \, R_{{\rm in}}$ as compared to $R_{{\rm out}} = 200 \, R_{{\rm in}}$. Thus we used a fixed value of $R_{{\rm out}} = 200\, R_{{\rm in}}$ as a lower limit for the shell thickness, as this is computationally less expensive.

 The outflow dynamics, and in particular the density and velocity of the wind, play an important role in the emergent spectra. Typical wind velocities derived from observations of Galactic AGB stars range from 10 to 20 kms$^{-1}$ \citep{Vassiliadis1993, Bloecker1995, HabingBook}. In the lower-metallicity environments of the LMC and SMC it is expected that the wind speed in O-AGB stars is smaller compared to their Galactic counterparts \citep{Marshall2004}. We adopt a constant outflow velocity for the dust of $v_{{\rm exp}} = 10$ kms$^{-1}$  for all the Magellanic Cloud object \citep{Wood1992, vanLoon2001}. For the assumptions adopted in the present paper, the mass-loss rate is linearly related to the actual individual expansion velocity via $\dot{M}$ $\propto$ $ v_{{\rm exp}}/(10 \, {\rm km \, s}^{-1})$.

Total mass-loss rates are varied from 10$^{-10}$ to $10^{-5}$ M$_{\odot }$ yr$^{-1}$ and the dust-to-gas ratio is taken to be $\Psi$ = 0.005 which is typically used for O-AGB stars in the LMC \citep{Groenewegen2006, Sargent2011, Srinivasan2011}.

This enables us to encompass a large range of spectral morphologies, from `naked' (dust-free) stars to stars heavily enshrouded by circumstellar dust with a large infrared excess ($[3.8]-[8] > 0.8$). 
Our model grid extends to a colour of $[3.8]-[8] = 2.4$ which covers 95 per cent of the confirmed O-rich AGB stars in the LMC. Extreme O-rich stars that are redder than this limit have mass-loss rates greater than $10^{-5}$ M$_{\odot }$ yr$^{-1}$ \citep{Groenewegen2009}.

However, caution must be taken when comparing models with low dust-production rates ($\dot{D}$) to observations, as it is very difficult to differentiate stellar photospheres from sources with very low-contrast dust features below  $\dot{D}$ $\approx$ $10^{-11}$ M$_{\odot }$ yr$^{-1}$ \citep{Riebel2012}, since the dust-emission contrast is proportional to the dust-production rate.

For all our models a distance of $ 49.97 \, \rm {kpc} $ to the LMC is assumed \citep{Pietrzynski2013}.   


\subsubsection{Dust grain properties}

\begin{figure*}
\includegraphics[trim=0cm 0cm 0cm 0cm, clip=true, width=0.975\textwidth]{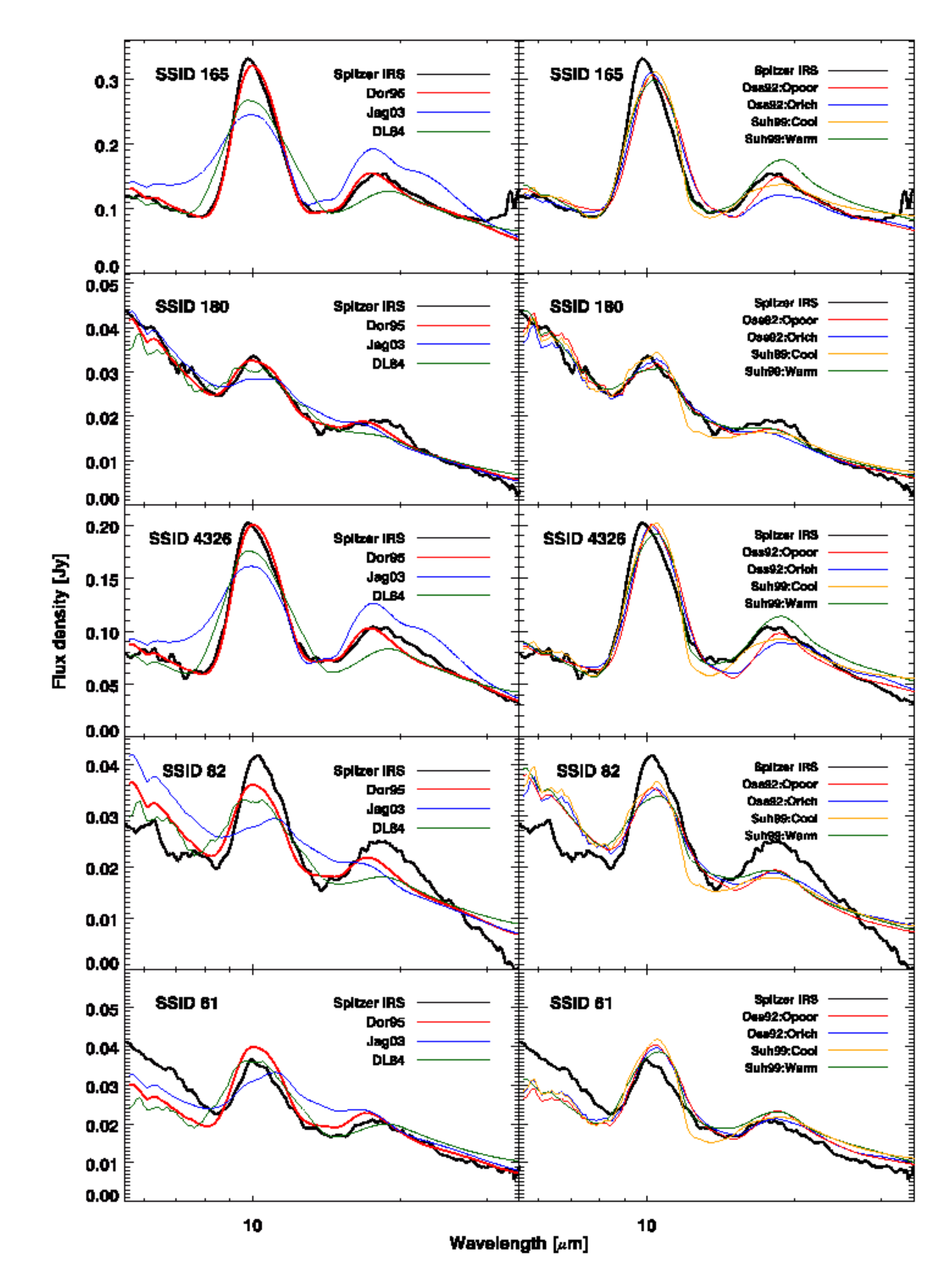}
 \caption{Comparison of astronomical and laboratory silicates to  {\em Spitzer} IRS observations of O-rich AGB stars in the LMC. The optical constants derived by Dorschner et al.~1995 provide the best overall fit to the spectra.} 
 \label{Fig:optCcomp}
\end{figure*}

The circumstellar envelopes of O-AGB stars are mineralogically complex. For stars with low mass-loss rates ($\dot{M} < 10^{-7} $M$_{\odot }$ yr$^{-1}$) simple metal oxides are the most abundant species \citep{Cami2002,Posch2002}; at greater mass-loss rates silicates (in both amorphous and crystalline form) become the dominant dust component. Metallic iron grains are thought to also contribute to the infrared emission around oxygen-rich AGB stars \citep{Kemper2002, Verhoelst2009, McDonald2010}, and iron may be incorporated in other dust grains, for example in amorphous silicates \citep{GailSed1999}. 

The chemical composition of amorphous silicate grains is not well known, and previous studies of oxygen-rich AGB stars have considered a wide range of dust compositions with varying success.  
In Figure~\ref{Fig:optCcomp} we compare five sets of `astronomical silicates' and two dust compositions obtained from laboratory measurements to {\em Spitzer} IRS observations of O-rich AGB stars in the LMC. The astronomical silicates we consider are the oxygen-deficient (Opoor) and oxygen-rich (Orich) silicates from \cite{Ossenkopf1992}, the `warm' and `cool' silicates from \cite{Suh1999}, and the \cite[][hereafter DL84]{Draine1984} astronomical silicates derived from ISM lines-of-sight.  For the laboratory measured dust species we consider the amorphous olivine refractory indices from \cite{Jager2003} and \cite{Dorschner1995}.

We find that the pure amorphous olivine from \cite{Dorschner1995} best reproduce the position and relative strengths of the 10 $\mu $m and 20 $\mu $m features in the infrared spectrum of O-rich AGB stars in the LMC. We assume that the stoichiometric composition of the amorphous silicate dust grains is not dependent on the mass-loss rate of the AGB star.

To increase the opacity in the near-IR region, we also include metallic iron, using laboratory data from \cite{Ordal1988}. We adopt a metallic iron abundance of 4 per cent by mass with respect to the amorphous silicates, following \cite{Kemper2002, deVries2010}.

For the refractive indices of amorphous alumina dust, we use the optical constants for porous alumina grains measured by \cite{Begemann1997}, which are extended to shorter wavelengths ($\lambda < 7.8 \, \mu$m) by concatenation with optical constants from \cite{Koike1995}.

For all dust species modelled we adopt a standard Mathis-Rumpl-Nordsieck (MRN) grain size distribution \citep{Mathis1977}, given by $N(a) \propto a^{-q}$, with $q$= 3.5, for a grain size range of $a = 0.01 - 1 \mu$m. 
To calculate the absorption and scattering coefficients of the dust grains from the complex refractive indices measured in the lab, an assumption about the grain shape distribution needs to be made; we adopt a continuous distribution of ellipsoids (CDE) for the particle shape \citep{BohrenHuffman1983}. This is preferred over homogeneous spherical grains as regularly-shaped particles introduce resonance effects resulting in unrealistic feature shapes \citep{Min2003}.

\begin{figure}
\includegraphics[width=84mm]{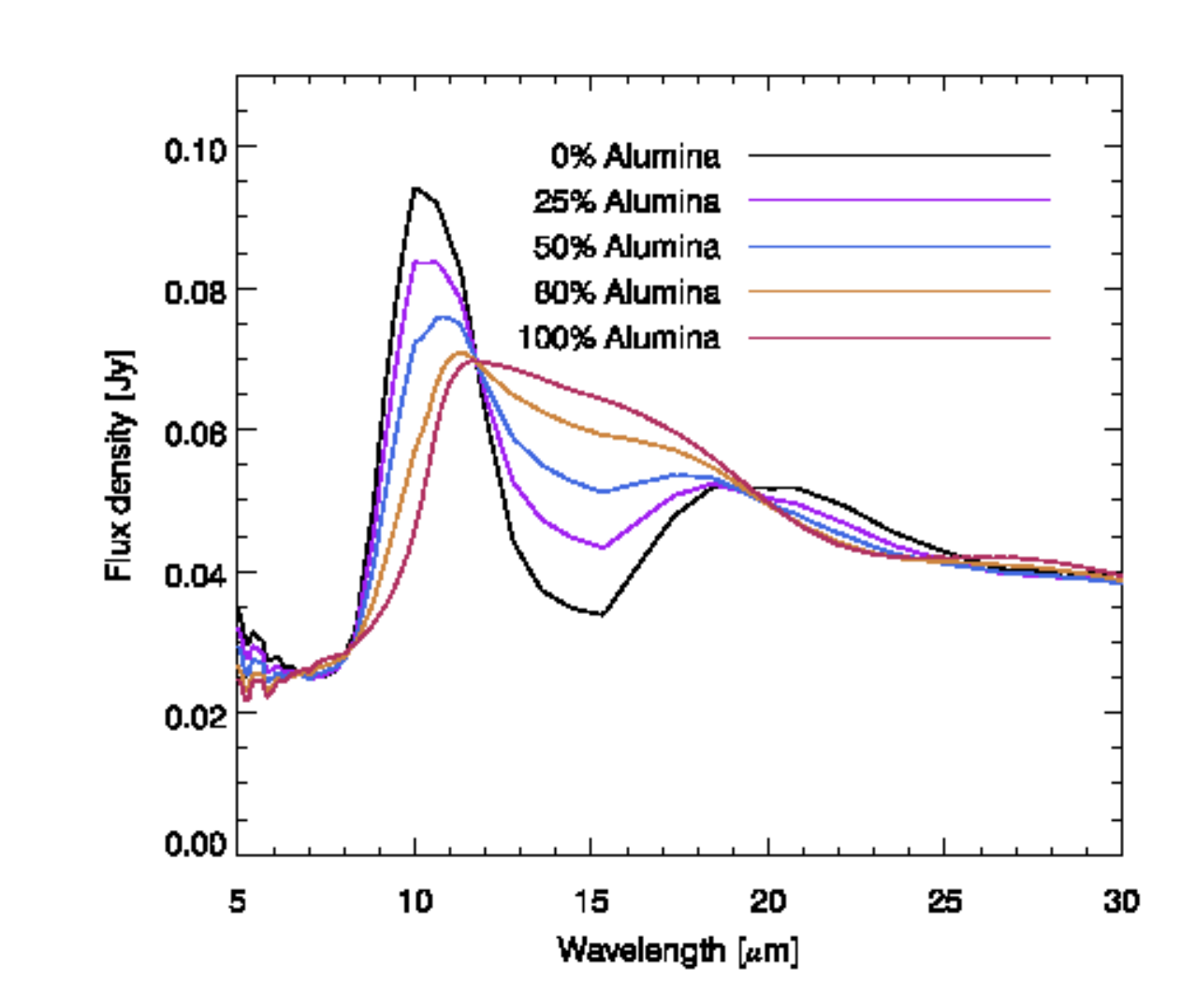}
 \caption{Five spectra with different alumina abundances are shown, for $\dot{M} = 7.5 \times 10^{-7}$ M$_{\odot }$ yr$^{-1}$. All other model parameters are constant.}
  \label{Fig:Alx_spec}
\end{figure}

The alumina abundance was varied in increments of 5 per cent for 0--40 per cent, and increased in steps of 10 per cent for 40--100 per cent. Spectra with different alumina abundances are shown in Figure~\ref{Fig:Alx_spec}; there is a clear change in the shape and strength of the 10- $\mu$m feature as the alumina abundance is increased.


\section{Model results} \label{results}

From our models we compute expected flux densities for the broad-band filters of 2MASS, {\em Spitzer}, {\em WISE} and {\em AKARI} by convolving the {\sc modust} spectral output with the relative spectral response curves. The calculated fluxes are further converted into Vega magnitudes, which can be directly compared to catalogue values.

\subsection{Colour--colour diagrams}\label{sec:ALgrid_CCDs}

\begin{figure*}
\centering
\includegraphics[width=84mm]{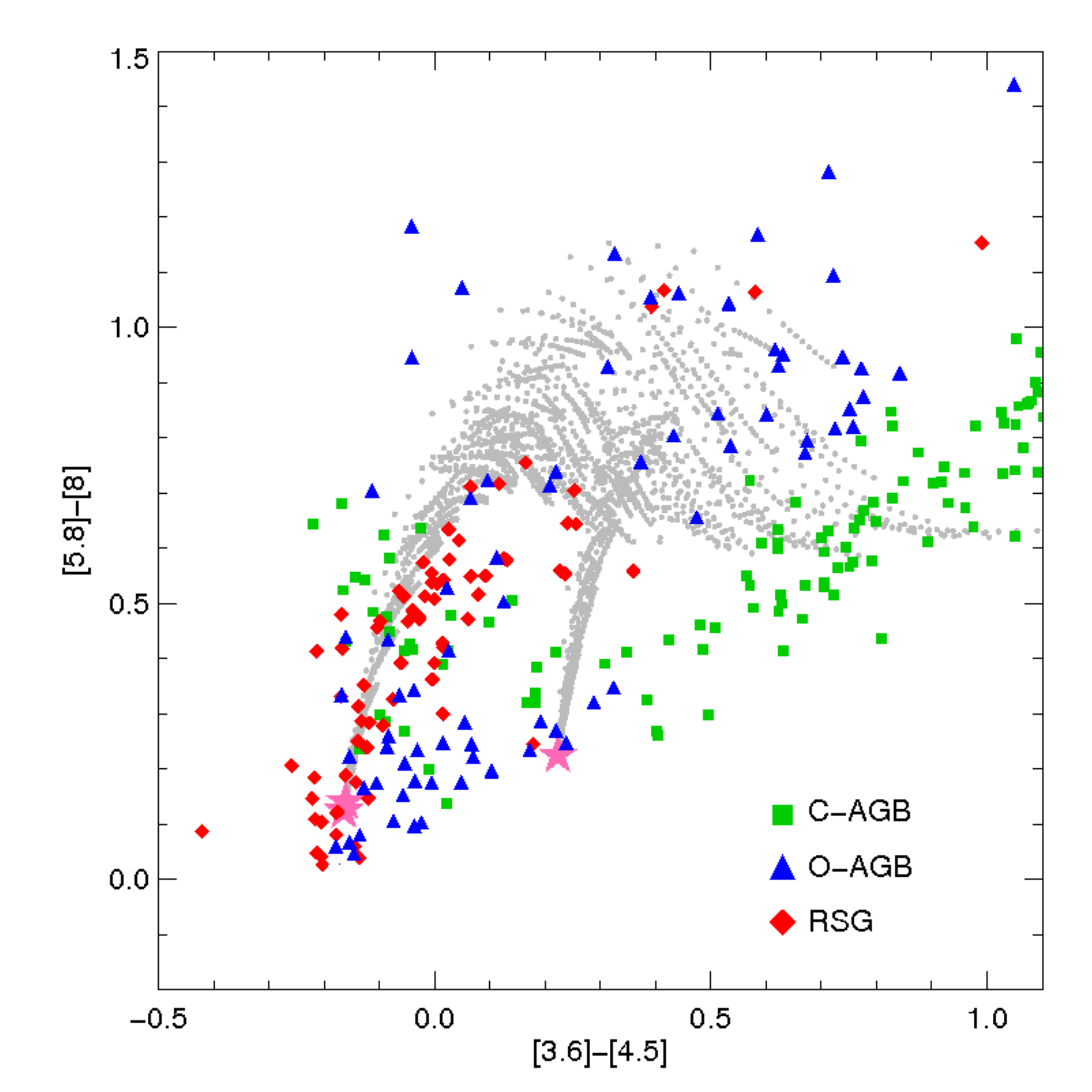}
\includegraphics[width=84mm]{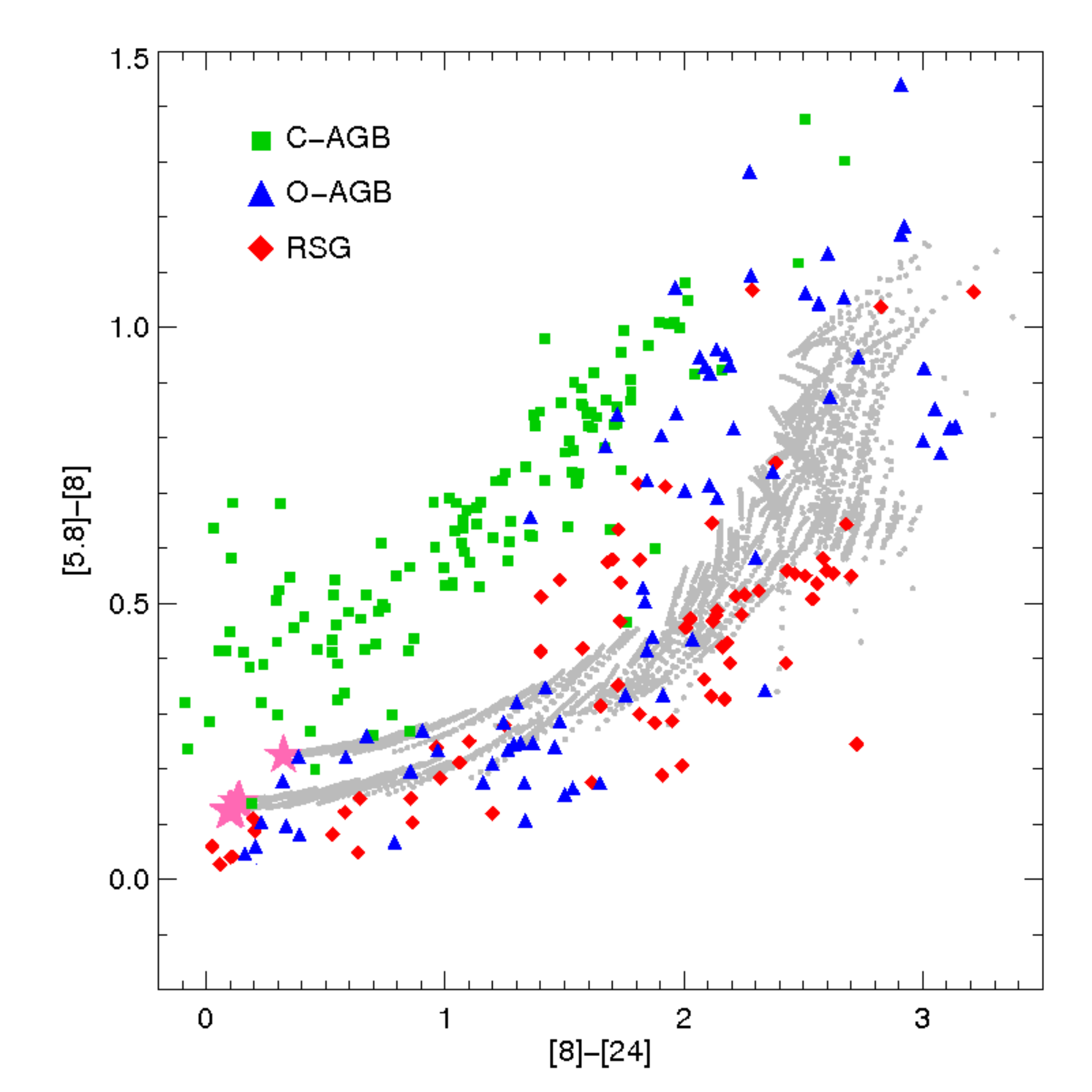}
 \caption[IRAC/MIPS colour-colour diagrams illustrating model grid coverage]{{\em Spitzer} IRAC/MIPS colour-colour diagrams for two combinations of colours. Small, grey dots denote the  models from our O-rich grid.  Also shown are the \cite{Fluks1994} photospheres (pink stars) used to generate the grid. To illustrate the model grid coverage, the evolved sources in the SAGE-Spec sample are overlaid: O-rich AGB stars are represented by  blue triangles, RSGs by red diamonds and C-rich AGB stars by green squares.}
  \label{fig:Al_grid_CCDs}
\end{figure*}

In this section we show examples of the colour--colour  space occupied by our grid of models and compare our synthesised photometry to observations of evolved stars. Since our models are created for dusty O-rich AGB stars we focus on mid-IR colours, where molecular and dust spectral features cause distinct photometric signatures and the central star's effective stellar temperature has little influence on the flux.

\begin{figure*}
\centering
\includegraphics[width=84mm]{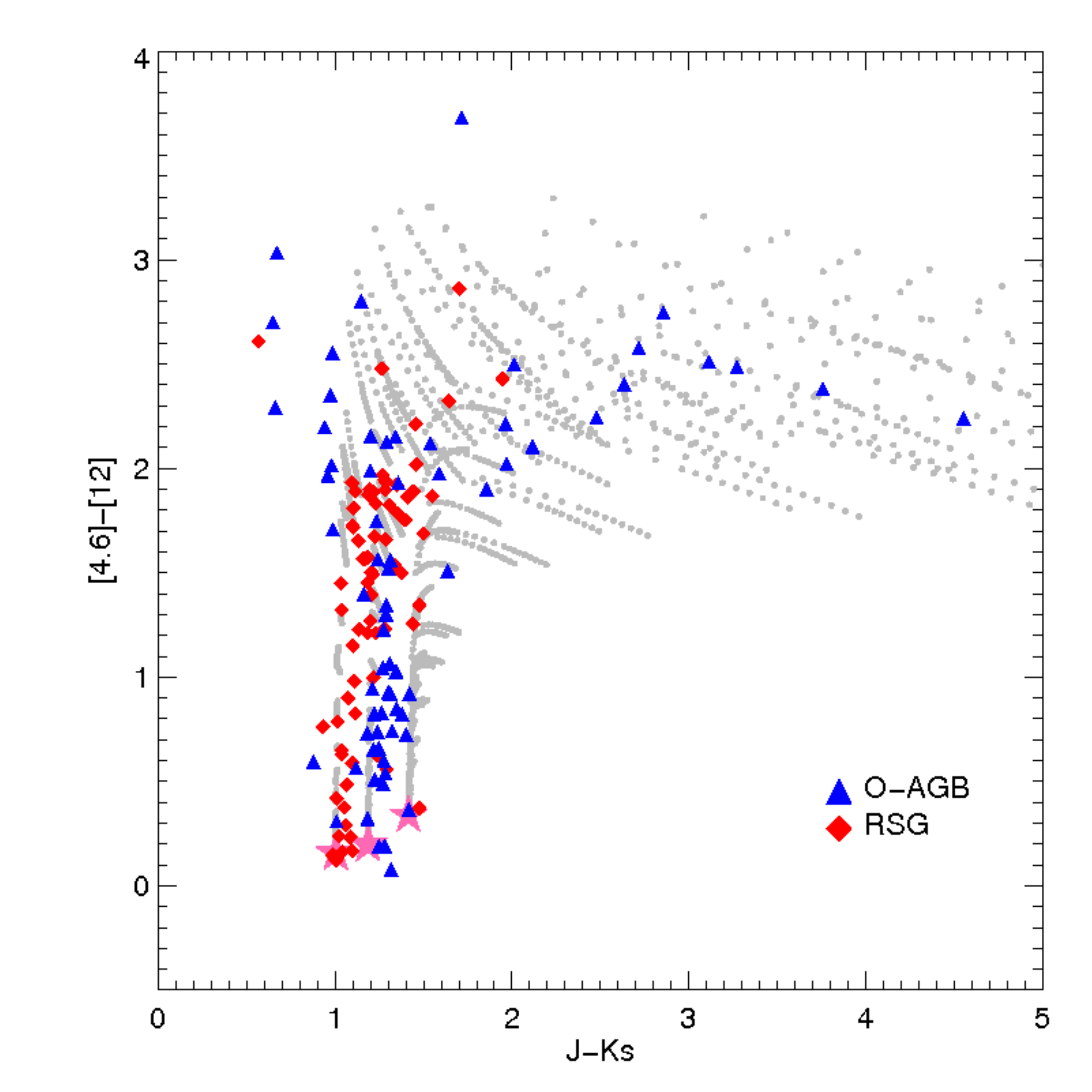}
\includegraphics[width=84mm]{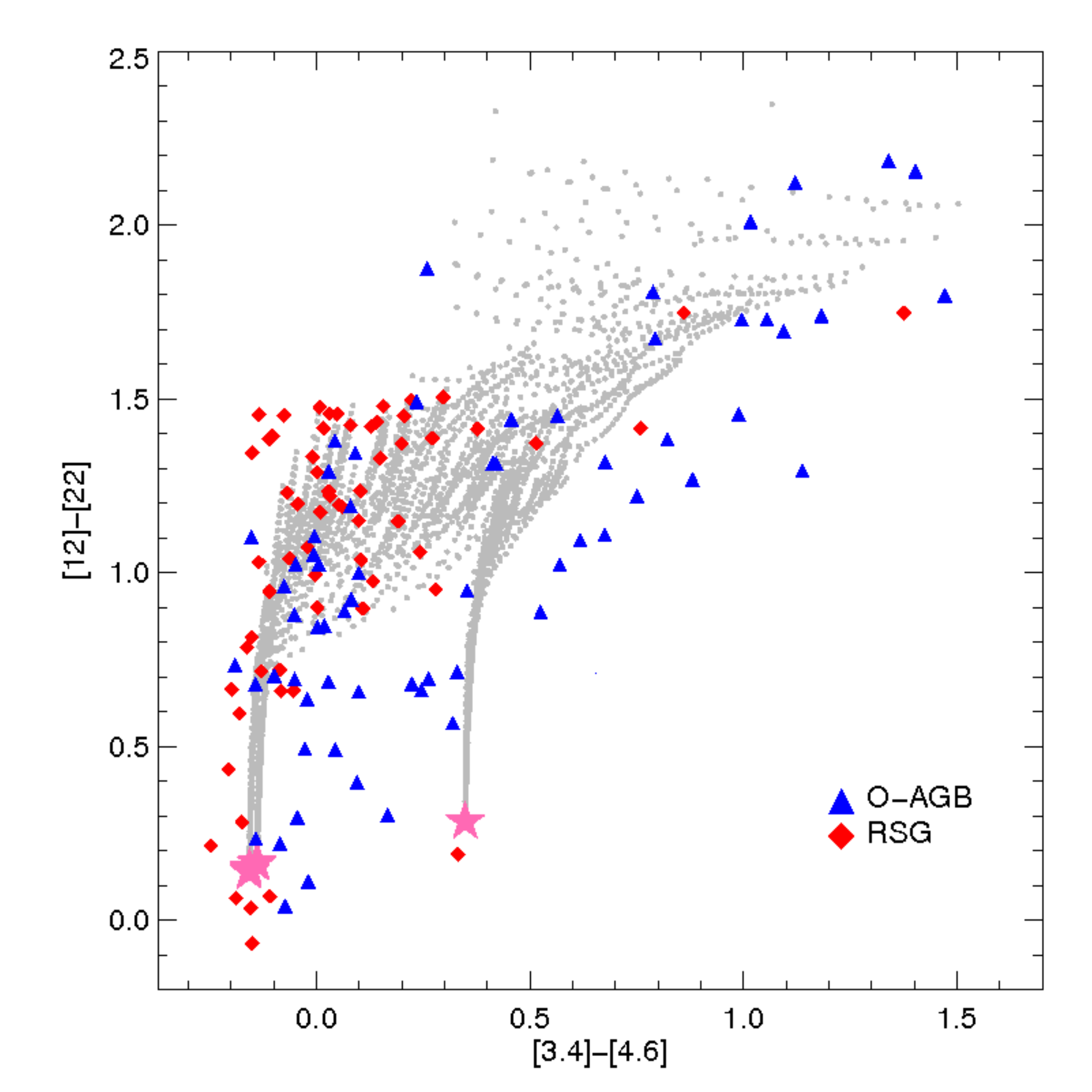}
\caption[2MASS/{\em WISE} colour-colour diagrams illustrating model grid coverage]{2MASS--{\em WISE} J-K$_{\rm{s}}$ vs.~$[4.6]-[12]$ colour-colour diagram (left) and {\em WISE} $[12]-[22] $ vs.~$[3.4]-[4.6]$ colour-colour diagram (right) of the known O-AGB and RSG stars in the SAGE-Spec sample. Symbols  are as in Figure \ref{fig:Al_grid_CCDs}.}
\label{fig:Al_grid_WISE_CCD}
\end{figure*}

\begin{figure*}
\centering
\includegraphics[width=84mm]{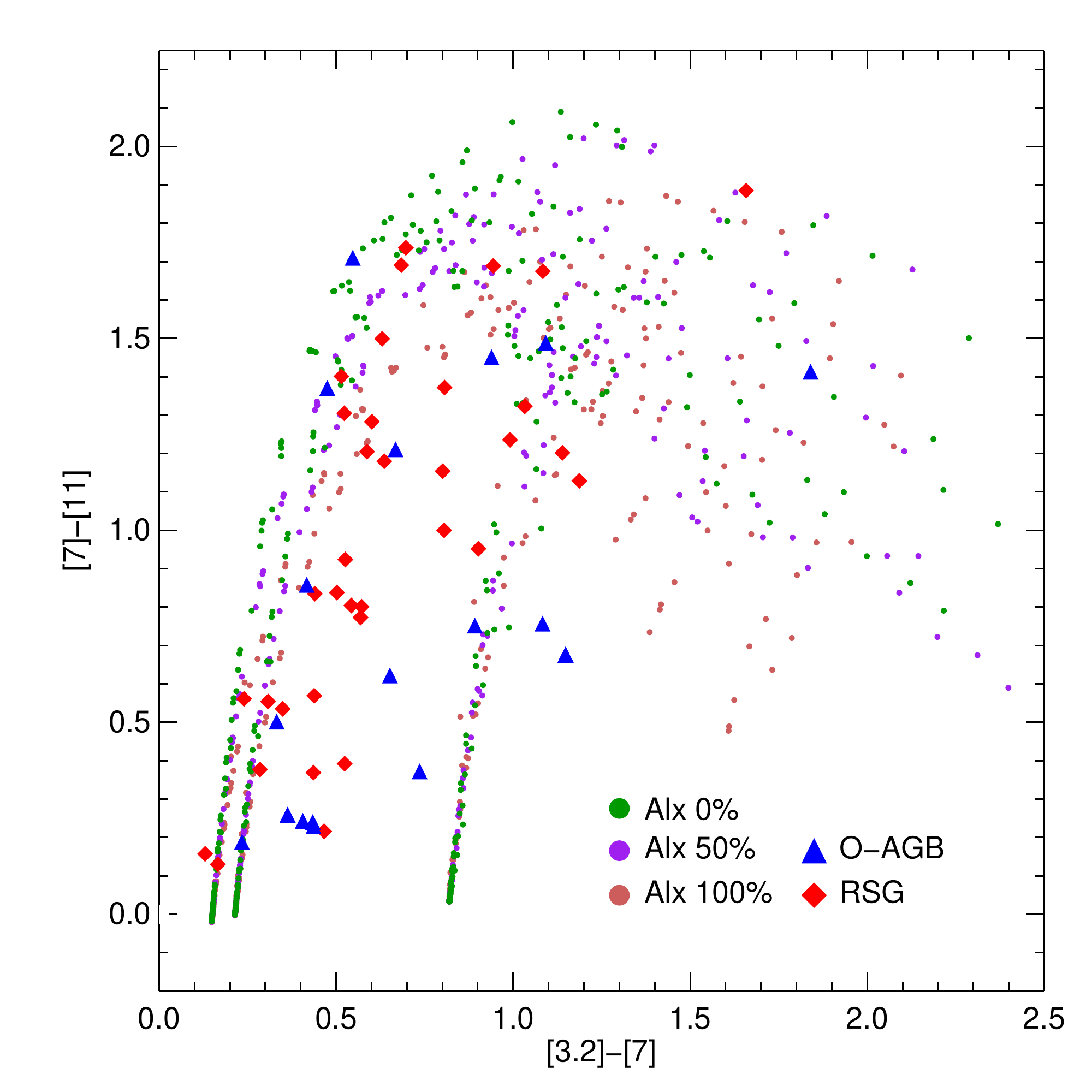}
\includegraphics[width=84mm]{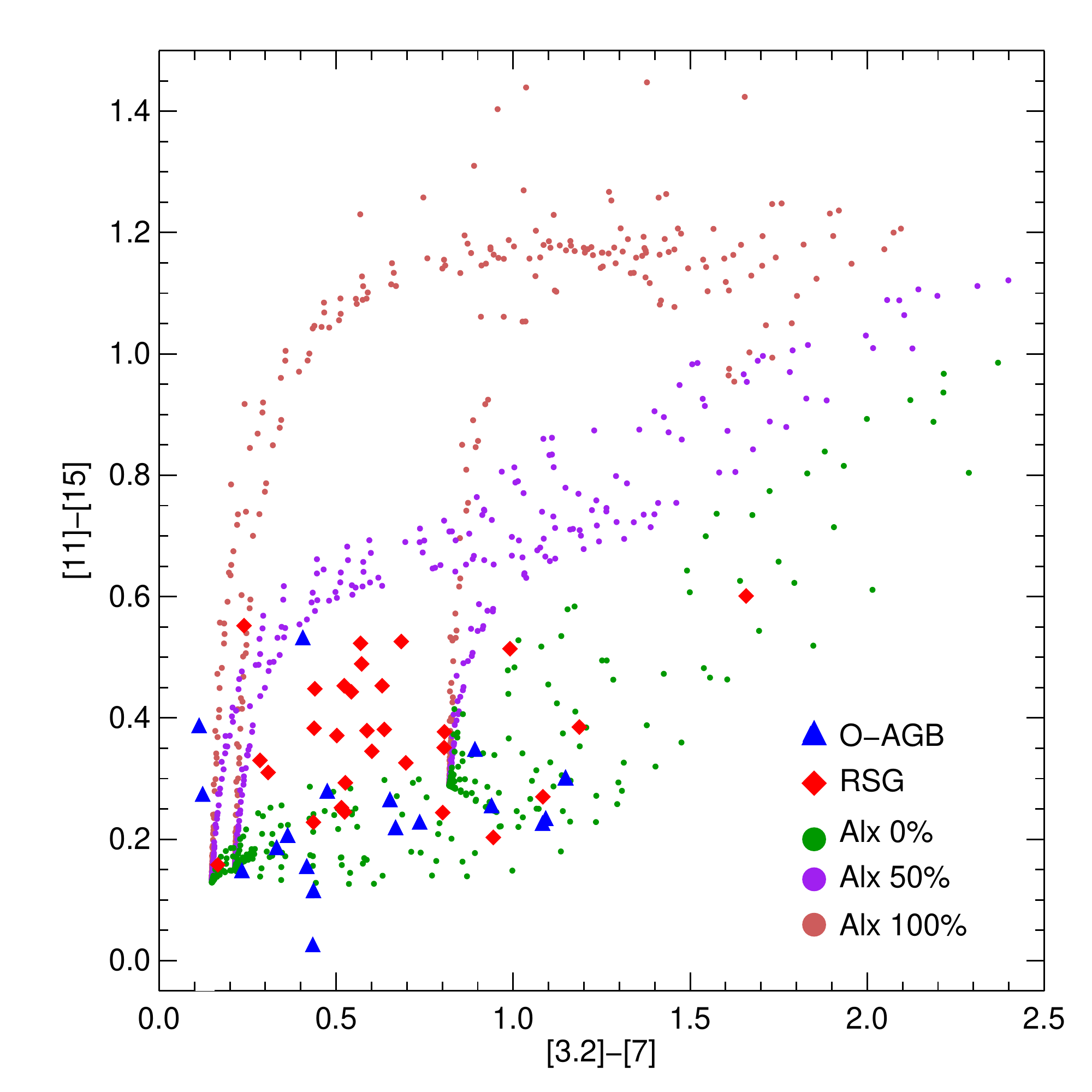}
 \caption[AKARI colour-colour diagrams illustrating model grid coverage]{{\em AKARI} colour-colour diagrams for two combinations of colours. Models with pure amorphous silicates (green), pure alumina (brown) and an equal blend between alumina and silicates (purple) are highlighted. All other symbols are as in Figure \ref{fig:Al_grid_CCDs}.}
  \label{fig:Al_grid_AKARI_CCDs}
\end{figure*}

\begin{figure*}
\centering
\includegraphics[width=84mm]{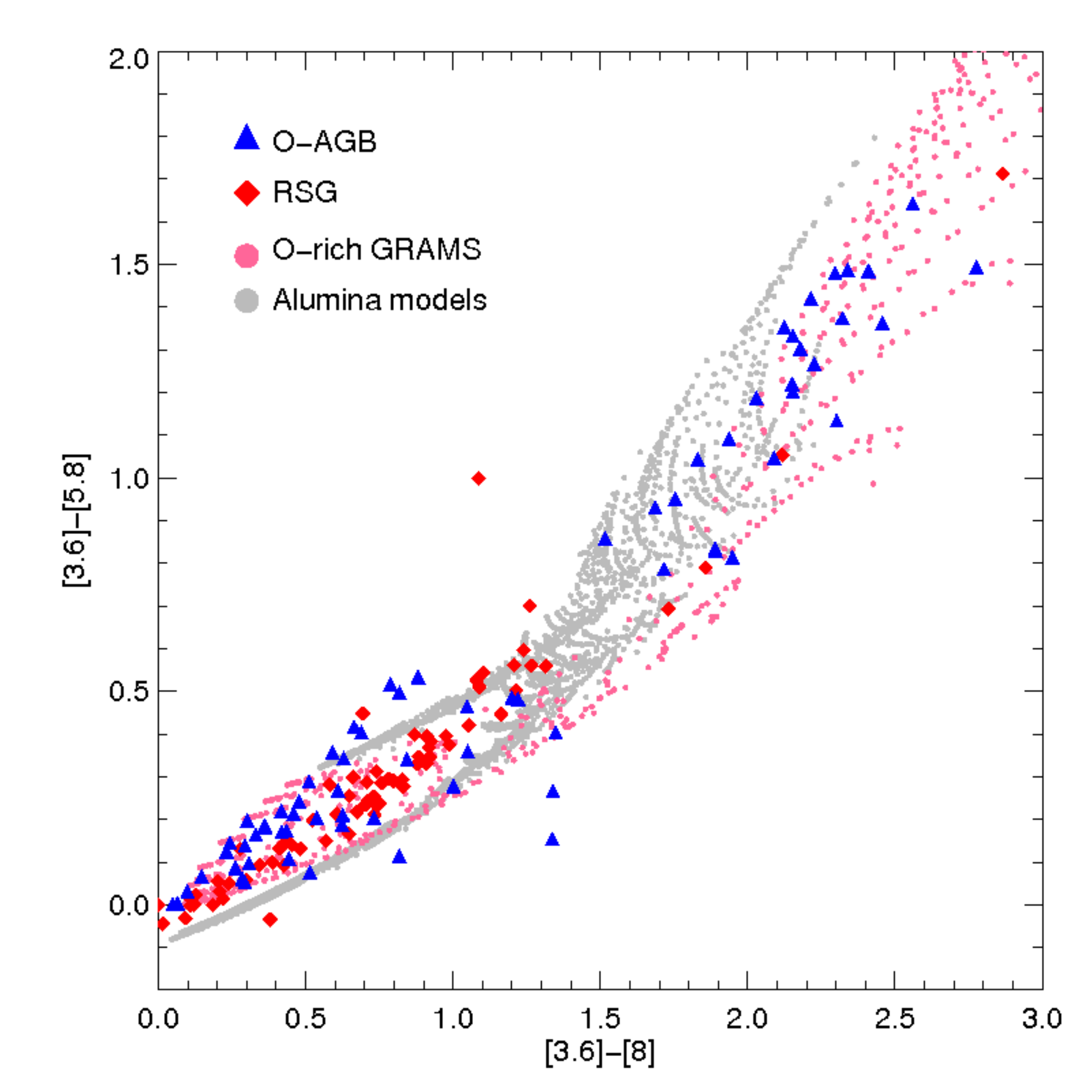}
\includegraphics[width=84mm]{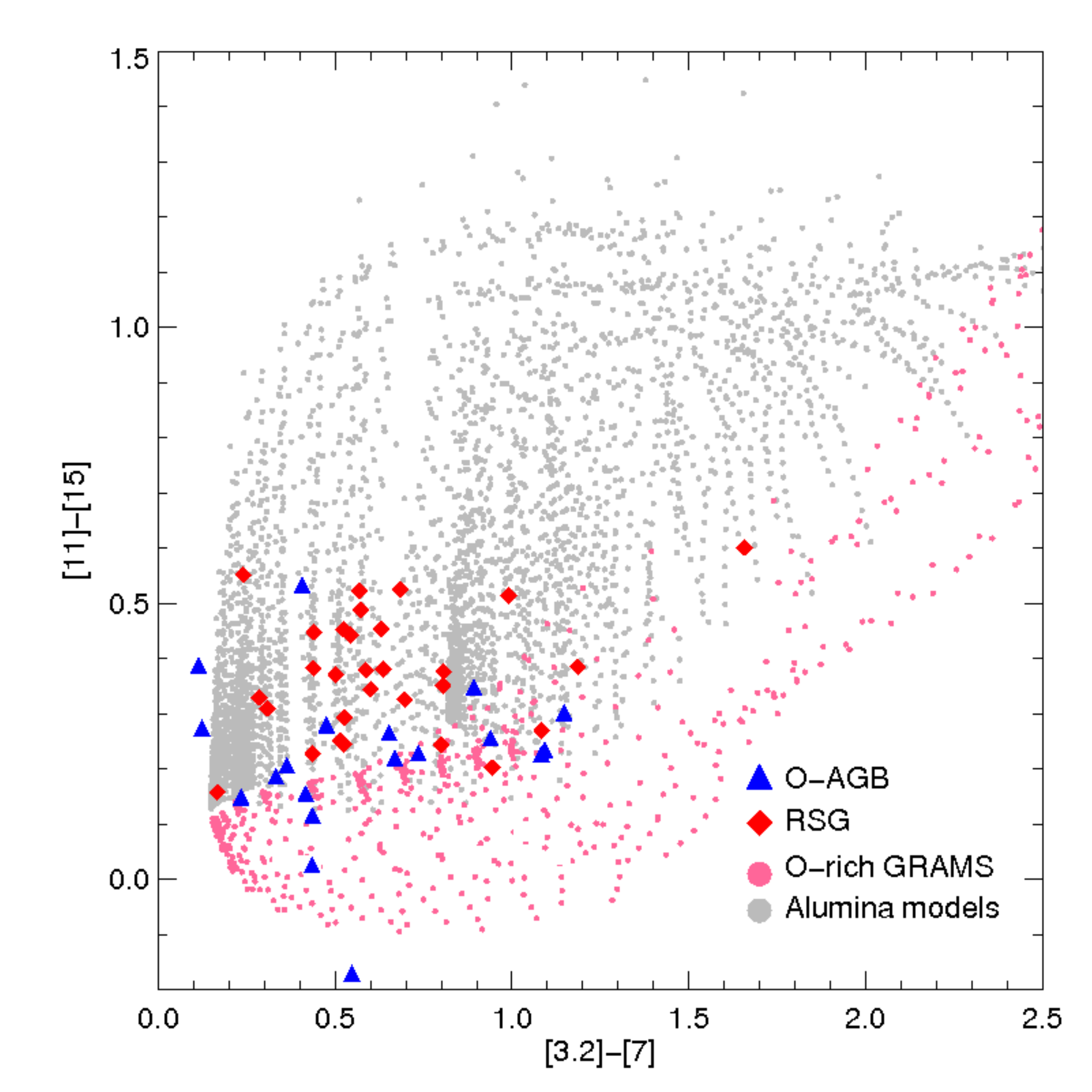}
 \caption[Comparison with the {\sc Grams} model grid]{Comparison of the coverage of the model grid described in this work with the oxygen-rich {\sc grams} models (pink points).}
  \label{fig:CCD_grids_comp}
\end{figure*}

Mid-IR colour--colour diagrams  (CCD) for the model grid are presented in Figures~\ref{fig:Al_grid_CCDs}--\ref{fig:Al_grid_AKARI_CCDs}. For comparison, we also include LMC point sources from the SAGE-Spec survey, which were identified as O-rich AGB stars and RSGs using {\em Spitzer} IRS spectra and ancillary photometry \citep{Kemper2010, Woods2010, Jones2012}. These were cross-identified with the {\em AKARI} LMC point source catalogue \citep{Ita2008, Kato2012} and the {\em WISE} all-sky catalogue \citep{Wright2010}.

In general, our models are consistent with the range of observed colours for oxygen-rich stars in the colour-colour diagrams considered here, especially when the contributions from circumstellar dust becomes significant in describing the observed photometry. However, the resolution of the grid before dust production begins is limited to narrow columns in colour-colour space (Figure~\ref{fig:Al_grid_WISE_CCD}), corresponding to sets of models with low optical depths derived from the three input stellar photospheres. The sparse grid coverage for the dust-free sources reflects the compromise between the accuracy of the model fit and the required computational time, with emphasis placed on the fine sampling of model AGB stars with apparent dust excesses.



The IRAC [5.8]$-$[8.0] versus [3.6]$-$[4.5] colour-colour diagram, shown in the left panel of Figure~\ref{fig:Al_grid_CCDs}, traces the photospheric temperature of the central star and continuum emission from the circumstellar dust shell. At these wavelengths O-rich dust has no distinguishable signatures, unless it contains a significant fraction of iron or has a strong 10-\mum silicate feature, both of which affect the 8-\mum flux. The IRAC colours are insufficient to distinguish between carbon-rich AGB stars, oxygen-rich AGB stars and RSGs, as such there is a significant overlap between the carbon-rich dust sources and our oxygen-rich models. 
The largest discrepancy between our model grid and observations is for sources that lie in the largely-dustless region below [5.8]$-$[8.0] $< $ 0.6 mag. These sources have little circumstellar excess, and may be better represented by photospheres with a larger range of stellar effective temperatures, metallicity and surface gravity. With the onset of mass-loss, the dust emission increases and the model colours provide better coverage of the observed data. For sources with a dust excess discrepancies between the models and observations may be accounted for by altering the metallic iron abundance, which influences the [3.6]$-$[4.5] colour in O-rich stars.


Moving to longer wavelengths, the [5.8]$-$[8] versus [8]$-$[24] CCD (Figure~\ref{fig:Al_grid_CCDs}; right panel) is an effective tracer of stars with significant amounts of circumstellar dust. It is often used to distinguish oxygen-rich from carbon-rich stars as there is little overlap between the populations \citep{Kastner2008,Boyer2012}. Our models on the whole show a considerable overlap with the colours of the O-rich AGB stars from SAGE-Spec sample but are well separated from the region occupied by the C-rich AGB stars.


The left panel in Figure~\ref{fig:Al_grid_WISE_CCD} shows a colour-colour diagram incorporating both near- and mid-IR data. This essentially compares the model photospheres via the J$-$K$_{\rm s}$ colour to the dust excess, traced by the  {\em WISE} [4.6]$-$[12] colour. In the diagram the different stellar types are reasonably well separated (due to differences between the stars effective temperature), until the stars become heavily dust-enshrouded at [4.6]$-$[12] $>$ 1.6. 
The observed RSGs, with bluer (J$-$K$_{\rm s}$) colours, are generally traced by models with an M1 photosphere, while the O-AGB stars that have slightly redder J$-$K$_{\rm s}$ colours are represented by the cooler M5 and M9 photospheres. 
The models reproduce the range of observed colours quite well, except for sources which lie off to the left of the model grid. This discrepancy is most likely due to an enhanced scattering by dust in an asymmetric shell or due to pulsations which can significantly effect the near-IR flux of AGB stars \citep{Whitelock2003}.
For Mira variable stars pulsation amplitudes are typically in the region of 6-8 magnitudes in V and are usually less than 2.5 mag in K. RSG are typically less variable than AGB stars; this class of objects has a lower fraction of sources which diverge from the model grid, in this case scattering of light in an asymmetric dust shell may be important.

The {\em WISE} [12]$-$[22] versus [3.4]$-$[4.6] CCD shown in the right panel  of Figure~\ref{fig:Al_grid_WISE_CCD} has a similar model coverage to the {\em Spitzer} [5.8]$-$[8.0] versus [3.6]$-$[4.5] CCD. The {\em WISE} [3.4]$-$[4.6] colour is comparable to the {\em Spitzer} IRAC [3.6]$-$[4.5] colour and is a good indicator of the emission from the warmest regions of the dust shell, while the 12- and 22-$\mu$m filters measure the emission from the silicate features at 10 and 20 $\mu$m and thus this colour is a good tracer of oxygen-rich dust. As before a number of the O-rich AGB and RSG stars which contain little or no dust around them are not covered by the model grid. Additionally, some sources with [3.4]$-$[4.6] $>$ 0.5 mag and  [12]$-$[22] $<$ 1.5 mag fall outside the region covered by our model grid; these sources may be better represented by models with a higher percentage of metallic iron grains in the circumstellar shell, which would increase the  [3.4]$-$[4.6] colour.


Figure~\ref{fig:Al_grid_AKARI_CCDs} shows two-colour diagrams using {\em AKARI} colours of [7]$-$[11] versus [3.2]$-$[7] and [11]$-$[15] versus [3.2]$-$[7]. The models with pure amorphous silicates, pure alumina and a equal blend between alumina and silicates are highlighted. While there is considerable degeneracy in the [7]$-$[11] colour, which measures the  10-\mum feature irrespective of the Al$_{2}$O$_{3}$ content,  the [11]$-$[15] colour provides a measurement of the change in strength/shape of the 10-\mum dust feature.  This diagram separates the different dust types well, in sources with a reasonable dust excess. It can therefore be used as a diagnostic to derive the fractional abundance of alumina (Alx). The sources in the SAGE-Spec sample occupy a region in colour-colour space corresponding to an alumina fraction of less than 50 per cent (we return to this in Section~\ref{AluminaLMC}).

\begin{figure}
\centering
\includegraphics[trim=0cm 0cm 0cm 0cm, clip=true,width=84mm]{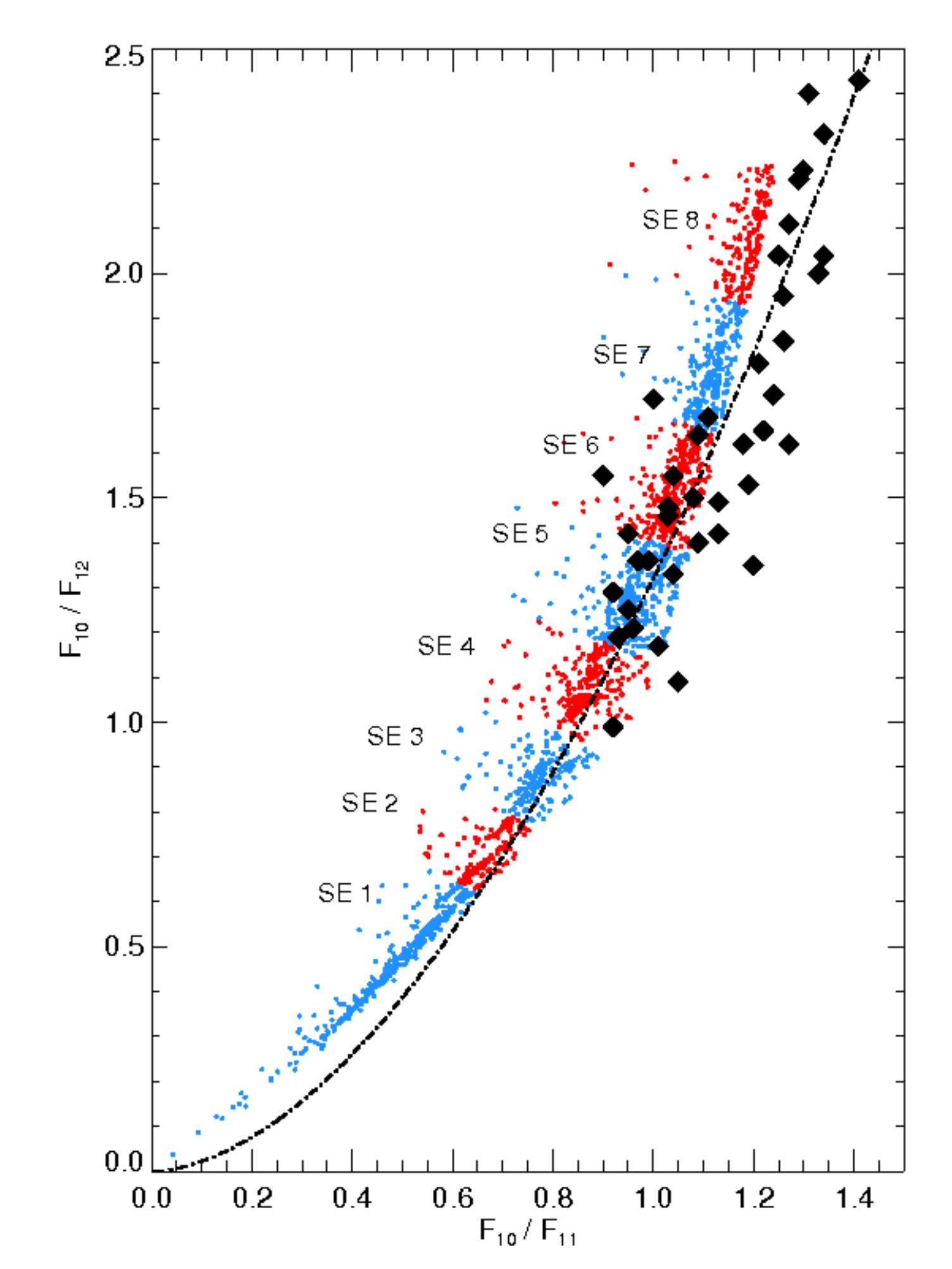}
 \caption{The silicate dust sequence power law: $F_{10}/ F_{12}=1.32( F_{10}/ F_{11}) ^{1.77}$ (dotted black line; from \citealt{Sloan1995}), and the flux ratios for the individual models. For comparison the filled black diamonds show the flux ratios of LMC O-AGB sources.}
  \label{fig:AlModelSEidx_seq1}
\end{figure}

\begin{figure}
\centering
\includegraphics[trim=0cm 0cm 0cm 0cm, clip=true, width=84mm]{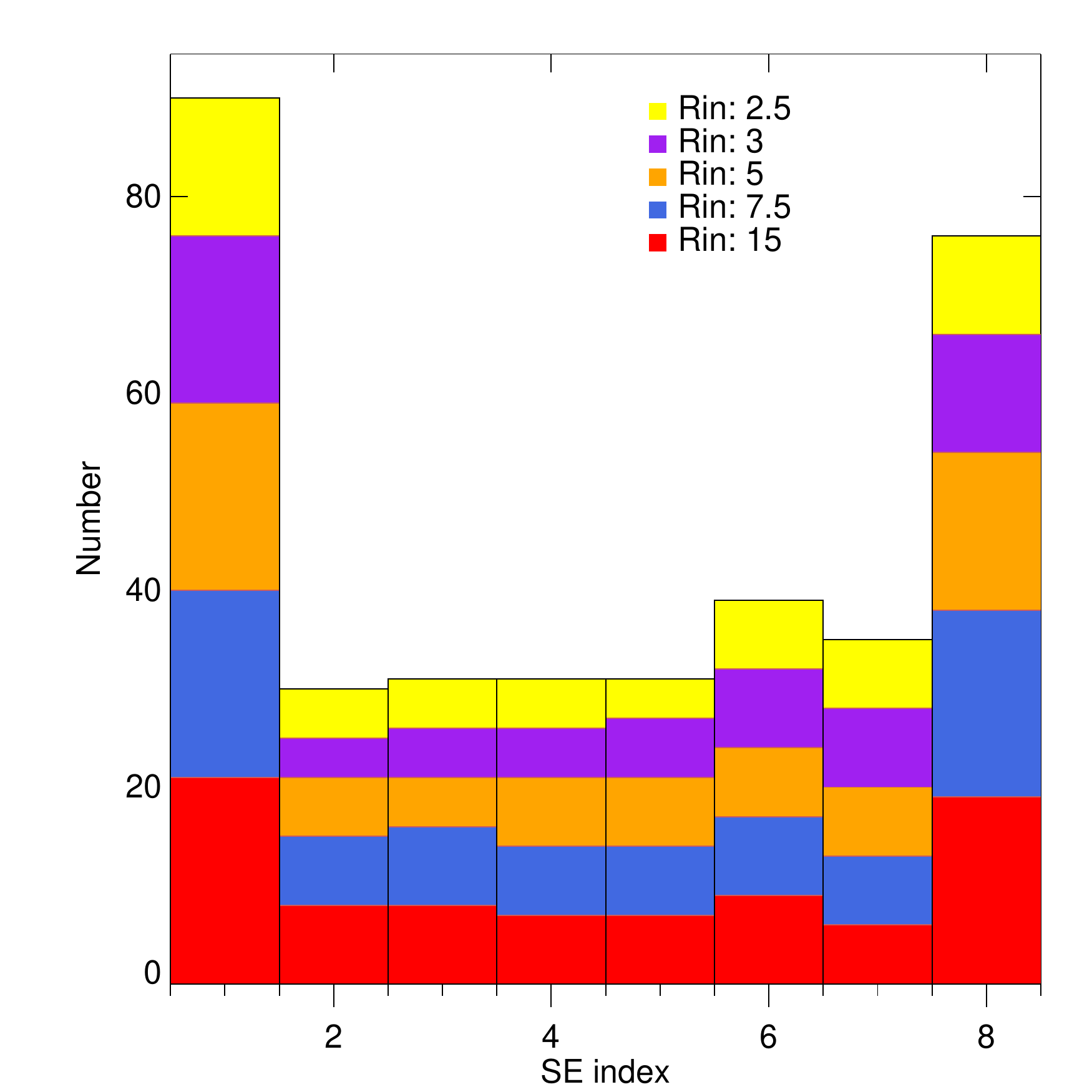}
 \caption{Distribution of the model grid silicate emission index. Each colour represents a dust shell inner radius explored by the grid for models with an M9 stellar photosphere. The even distribution between $R_{{\rm in}}$ models suggests that the SE index is independent of the dust temperature.}
  \label{fig:AlModelSEidx_hist}
\end{figure}

\subsubsection{Comparison to the {\sc grams} model grid}

In a recent similar study, \cite{Sargent2010} and \cite{Srinivasan2011} developed the Grid of RSG and AGB ModelS (GRAMS) to reproduce the range of observed infrared colours of LMC red supergiant (RSG) and AGB stars, and to measure their luminosities and dust production rates. The GRAMS grid consists of models for oxygen-rich as well as carbonaceous chemistries; the O-rich models \citep{Sargent2010} consist of oxygen-deficient silicate grains from \cite{Ossenkopf1992}, while the carbon-star dust \citep{Srinivasan2011} is composed of a mixture of amorphous carbon and silicon carbide, the latter making up 10 percent of the mass. The authors used the {\bf 2D}ust code  \citep{Ueta2003} to solve the radiative transfer problem for a dust shell of given inner and outer radius and optical depth around a central star that was represented by a model photosphere. They chose the range of input parameters (inner radii, optical depths) to cover the expected and observed values for LMC evolved stars.

We compare our model output to the  O-rich {\sc grams} models in Figure~\ref{fig:CCD_grids_comp}. Overall, there is good agreement between the sets of oxygen-rich models, particularly in the IRAC bands at shorter wavelengths ($\lambda < 8 \mu$m). In the near-IR the {\sc grams} grid agrees better with the bluer stellar sequences where stellar temperature has a greater influence on the colour.  Although both sets of model grids cover a similar range in stellar effective temperatures, the 14 {\sc phoenix} stellar photosphere models \citep{Kucinskas2005, Kucinskas2006} used in the O-rich GRAMS grid has a finer effective temperature resolution. The metallicity of the photospheres are also different, however for dusty stars this has a minimal effect on the broadband photometry and colour. In the mid-IR our models span a wider range of colour space due in part to the inclusion of alumina and the choice of optical constants for the amorphous silicates. 
It should also be noted that two models that overlap in colour-colour space may not necessarily have the exactly the same input parameters.

\subsection{SE index}  \label{sec:SEidx}

The silicate emission (SE) index is a spectral classification system developed by \cite{Sloan1995, Sloan1998} to measure the variation in the shape/strength of the emission feature at about 8--12 $\mu $m in  optically thin oxygen-rich AGB stars. The observed spectra are divided into eight categories (SE 1--8) based on the ratios of narrow-band fluxes at 10, 11 and 12 $\mu $m (F10, F11 and F12). These classes are designed to reflect the progression from the broad oxygen-rich dust emission features with little structure (SE 1--3) to the classic narrow 9.7-$\mu $m silicate features (SE 7--8).

The change in spectral features represents the dust formation process along the AGB, which depends strongly on $\dot{M}$ \citep{Dijkstra2005}. We would therefore expect the SE index to reflect the mass-loss rate, the chemical composition of the dust grains (i.e. the relative abundances of silicates and amorphous alumina), and variations in the temperature of the dust shell \citep{Ivezic1995,Hron1997,Egan2001}.

To investigate the parameter space covered by each SE class we apply the spectral classification procedure of \cite{Sloan2003} to each optically thin model in our grid.
To isolate the dust emission we subtract the appropriate stellar photosphere used in constructing the model. Each model's dust excess was quantified via the dust emission contrast (DEC), defined as the ratio of dust emission to stellar emission between 7.67 and 14.03 $\mu$m \citep{Sloan2008}. In keeping with observations, and to prevent any over-interpretation of models with no significant dust excess,  we do not ascribe an SE class to models with $\mathrm{DEC}\,\leq 0.10$. This corresponds to all models with a mass-loss rate below  $\dot{M}$ $\sim$ $3 \times 10^{-9}$ M$_{\odot }$ yr$^{-1}$.


Figure~\ref{fig:AlModelSEidx_seq1} shows the F10/F12 versus F10/F11 model flux ratios and the empirically derived silicate dust sequence power law: $F_{10}/ F_{12}=1.32( F_{10}/ F_{11}) ^{1.77}$ \citep{Sloan1995}, which forms the basis for the SE index classification.  Our models reproduce each of the eight SE indices along the silicate dust sequence; SE classes one and eight contain the largest numbers of models, with the remainder equally distributed across SE indices 2--7. This due to sources falling outside the SE1--8 range being classified as either SE1 or SE8.

There is a slight deviation from the silicate dust sequence power law at SE8. Observationally, sources in the Milky Way tend to occupy a region to the right of the silicate dust sequence for SE8, while stars from low-metallicity Galactic globular clusters populate a region to the left of the sequence \citep{Sloan2008,Sloan2010}. 
At the other end of the SE sequence, we suspect the curvature arises due to the models in our grid with high mass-loss rates combined with a high fractional abundance of alumina (Alx). 

The distribution of SE classes for our model grid is shown in Figure~\ref{fig:AlModelSEidx_hist}. Segregating the models according to the inner radius of the dust shell has no significant effect upon the SE index distribution, which suggests that the SE index is independent of the dust temperature. This is not surprising as the SE index is calculated using flux ratios across the 10 $\mu$m feature of a given temperature, effectively eliminating the grain temperature parameter. 
Since we have isolated the dust emission, the SE index is also independent of the spectral type and the effective temperature of the central star. 

\begin{figure*}
\includegraphics[trim=0cm 0cm 0cm 0cm, clip=true, angle=-90, width=\textwidth]{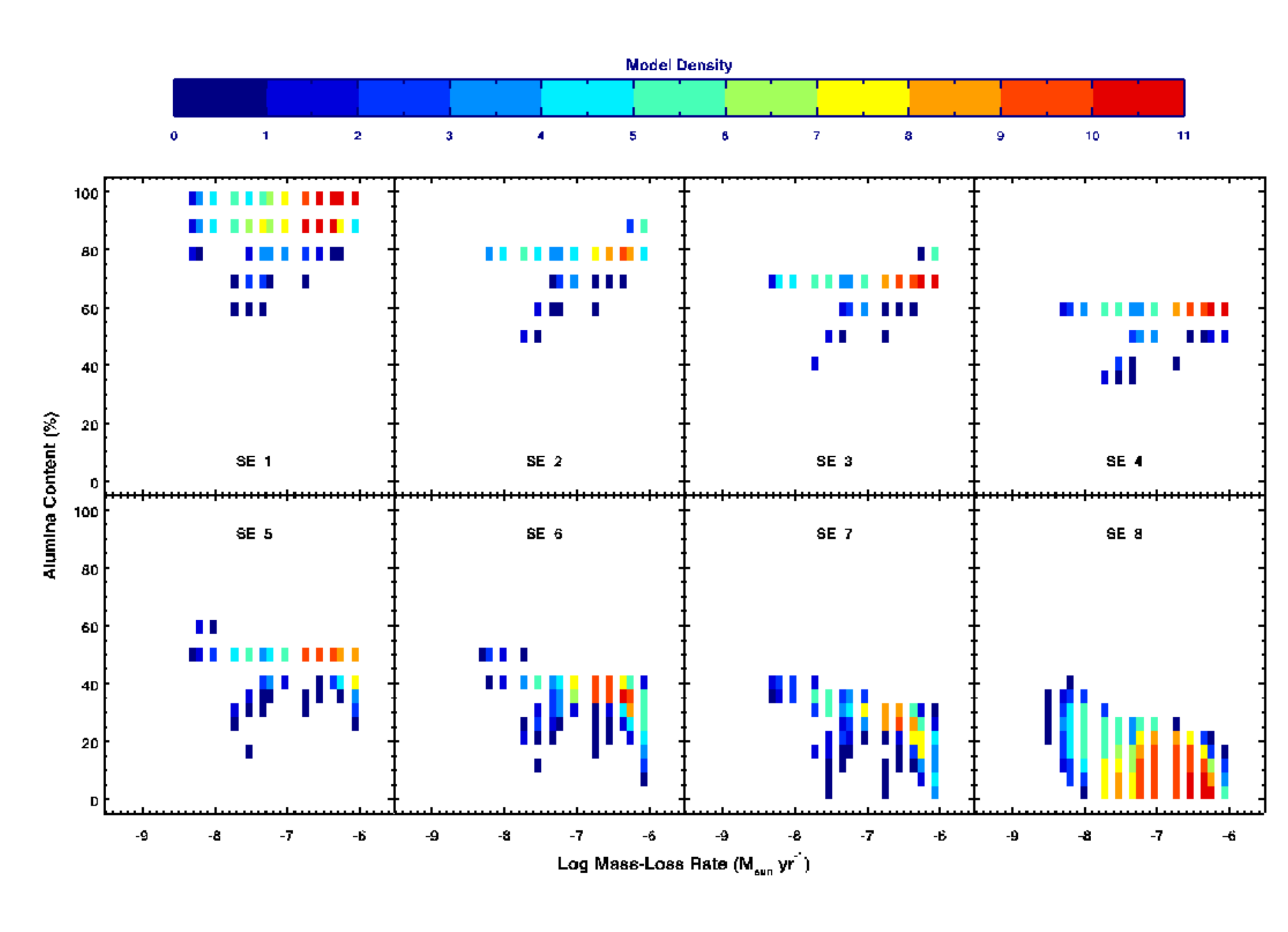}
 \caption{The relationship between the alumina content and total mass-loss rate for the models in our grid with a dust excess, separated according to the silicate emission index.}
  \label{fig:AlModelSEidx}
\end{figure*}

Figure~\ref{fig:AlModelSEidx} illustrates the range in mass-loss rate and alumina abundance covered by each SE class.  Increasing the density of the shell leads to models moving upward and to the right along the SE sequence, however, changes in position are generally to small to move a model to a higher SE class. For models dominated by either pure alumina or silicates, the shell density may have a greater influence as it determines when the emission from the dust shell becomes significant compared to that of the stellar photosphere.

The composition of the dust has a more substantial effect on the SE index than the mass-loss rate.  For a given mass-loss rate, a blend of alumina and silicates can reproduce the full silicate dust sequence. SE classes 1--3 are successfully reproduced by shells dominated by alumina dust (Alx $\gtrsim$ 70 per cent), while classes SE 6--8 are populated by models dominated by amorphous silicates (Alx $\lesssim$ 50 per cent). 

\begin{figure}
\centering
\includegraphics[trim=0cm 0cm 0cm 0cm, clip=true, width=84mm]{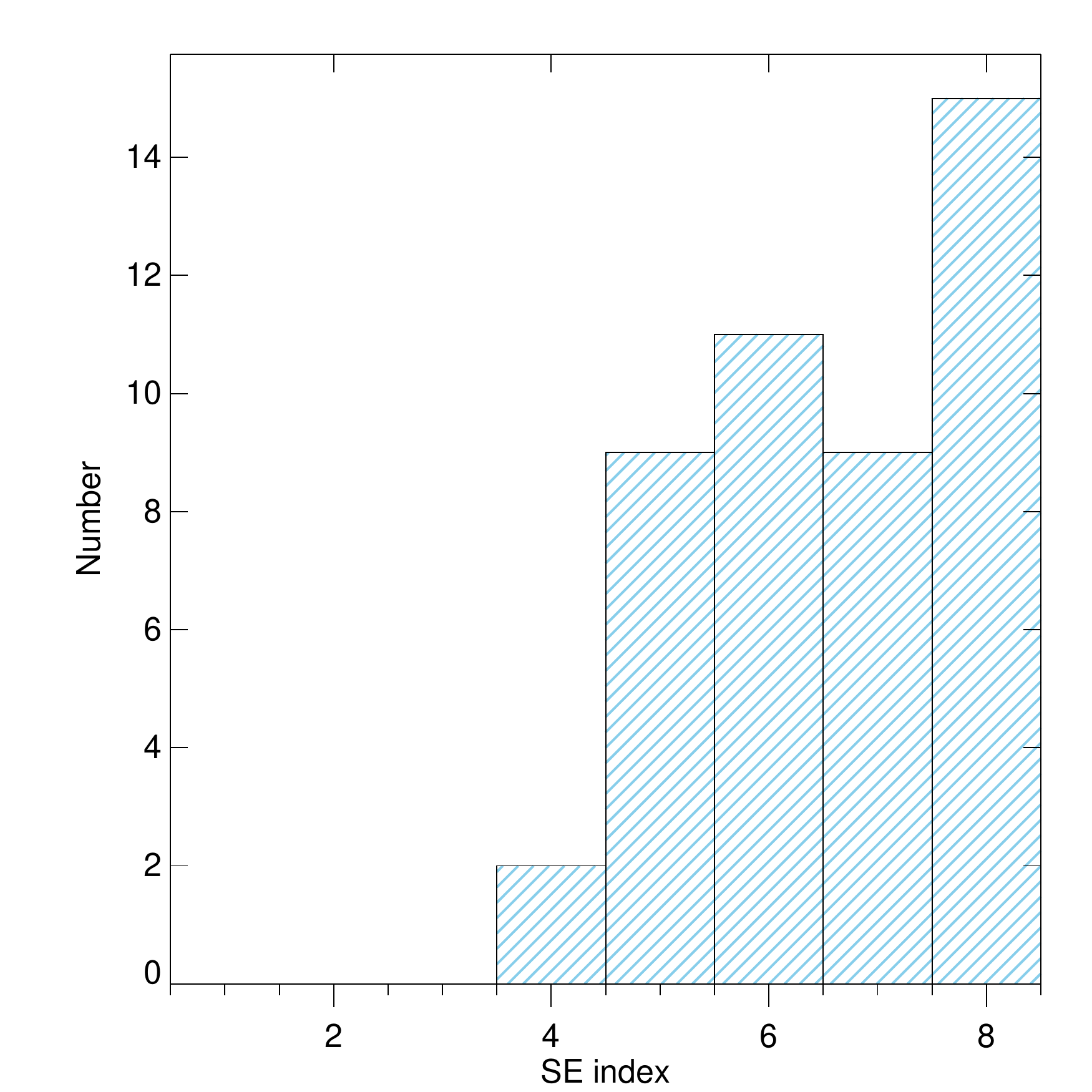}
 \caption[SE index distribution of O-AGB stars]{Distribution of the O-rich AGB stars in the LMC with silicate emission index.}
  \label{fig:LMCOAGB_SEidx_hist}
\end{figure}

The distribution of the O-rich AGB stars in the LMC with silicate emission index is shown in Figure~\ref{fig:LMCOAGB_SEidx_hist}. This suggests that the dust in the LMC is primarily silicate-rich, whilst alumina is largely absent, in agreement with the alumina dust abundances derived from {\em AKARI} photometric colours in Figure~\ref{fig:Al_grid_AKARI_CCDs}.  However, our spectroscopic sample only includes relatively bright AGB stars in the LMC, which have sufficient signal-to-noise in their IRS spectra to quantify the dust excess. Thus, the lack of sources with a low SE index could be due to an observational bias.

\section{Measuring the Alumina fraction}\label{AluminaLMC}

To estimate the alumina content of AGB stars in the LMC, we apply our model grid to the sample of 54 O-rich AGB stars described by \cite{Jones2012} taken from the {\em Spitzer} SAGE-Spec sample of \cite{Kemper2010} and the archival IRS spectra in the LMC. As alumina has not been observed in optically-thick AGB envelopes, we excluded the sources where the 10-$\mu$m feature is in self absorption from our analysis. Also excluded are sources that essentially appear as stellar photospheres with no significant infrared excess due to dust emission, but have been classified as AGB stars due to molecular absorption features and photometric variability. From this we are left with 37 sources. 


To model the AGB stars we have scaled the IRS spectrum to match the IRAC 5.8- and MIPS 24-$\mu$m fluxes to enable an accurate comparison to photometry. 
For each star we fitted the models to the observed broadband {\em JHK$_{\rm s}$}, IRAC and MIPS 24-$\mu$m photometric data and the flux measured over set wavelength intervals (line segments) from the spectra. This allows us to achieve a good fit to both the SED and the observed spectrum. The line segments have continuous wavelength coverage over the full {\em Spitzer} spectral range and each of the nine bands cover equal widths in logarithmic wavelength. This gives more weight to shorter wavelengths where the dust grains emit most of their flux and where the dust optical constants provide greater constraints to the observed emission feature profiles.
An additional six narrow bands (0.38 $\mu$m wide) over the 8--13 $\mu$m interval provide detailed measurements of the 10-$\mu$m feature.  
These 15 bands measure the spectral flux and place a strong constraint on the individual dust species present in the IR-spectrum. This is essential, as fits based only on the photometric data points poorly constrain the relative strengths of the dust features. 

Comparisons between the models and the data were made with an automated fitting routine and the best fit was selected based on a chi-square ($\chi ^2_\nu$) minimisation technique. The quality of the best-fit model was defined by:
\begin{equation}
\chi ^2_\nu = \frac{1}{N-p}\sum _i \frac{(f_{{\rm obs}_i} - f_{m_i})^2}{\sigma _i^2},
\end{equation}
where $f_{{\rm obs}_i}$ and $f_{m_i}$ are the observed and model flux in the ith band with error bar $\sigma _i$, $N$ is the total number of data points being fitted and $p$ is the number of free parameters in the model.
Uncertainties in the photometric data were modified to account for variability across a pulsation cycle \citep[see][]{Riebel2012,Jones2012}, and uncertainties in the spectroscopic data were inflated by a factor of five \citep[see][]{Groenewegen2009}.  

Our models are computed for a single luminosity, therefore they must be scaled to match the luminosity of the source in question before performing the $\chi^2$ fitting. The luminosity ($L$) of the source can be computed from this scaling factor ($\eta$) via $\eta = (L/7000 \, {\rm L}_{\odot})(d/49.97 \, {\rm kpc})^2$, where the distance ($d$) is assumed to be 49.97 kpc for the LMC.
As the luminosity is scaled, so too must the wind parameters be scaled appropriately. Following equation (2) of \cite{Groenewegen2006}, the stellar radius and the mass-loss rate must scale by the factor $\sqrt{\eta}$. The other model parameters and outputs are independent of this scaling.



Several models may provide a good fit to a source (defined arbitrarily by $\chi^2-\chi^2_{\rm min}<3$). 
To gauge the range a parameter could vary for a given source, we calculate its median absolute deviation; this is more resilient to outliers than the standard deviation and thus provides a more robust estimator of the uncertainty.
The SEDs and spectra for all the O-AGB stars in our sample  (identified by the SAGE-Spec designation; SSID) along with best-fitting model for each source are shown in Figure~\ref{fig:AlModelFit}.  To estimate a threshold value in the alumina abundance where its inclusion truly improves the fit, we also calculate the best-fit using models with a fixed oxygen-rich dust composition without an alumina component. We find that the adoption of amorphous alumina grains in the model significantly improves the fit when its abundance is greater than 15 per cent. The modelling results are summarised in Table~\ref{tab:AlmodelResults}. 


\begin{figure*}
\begin{center}
\includegraphics[trim=0cm 0cm 0cm 0cm, clip=true, width=\textwidth]{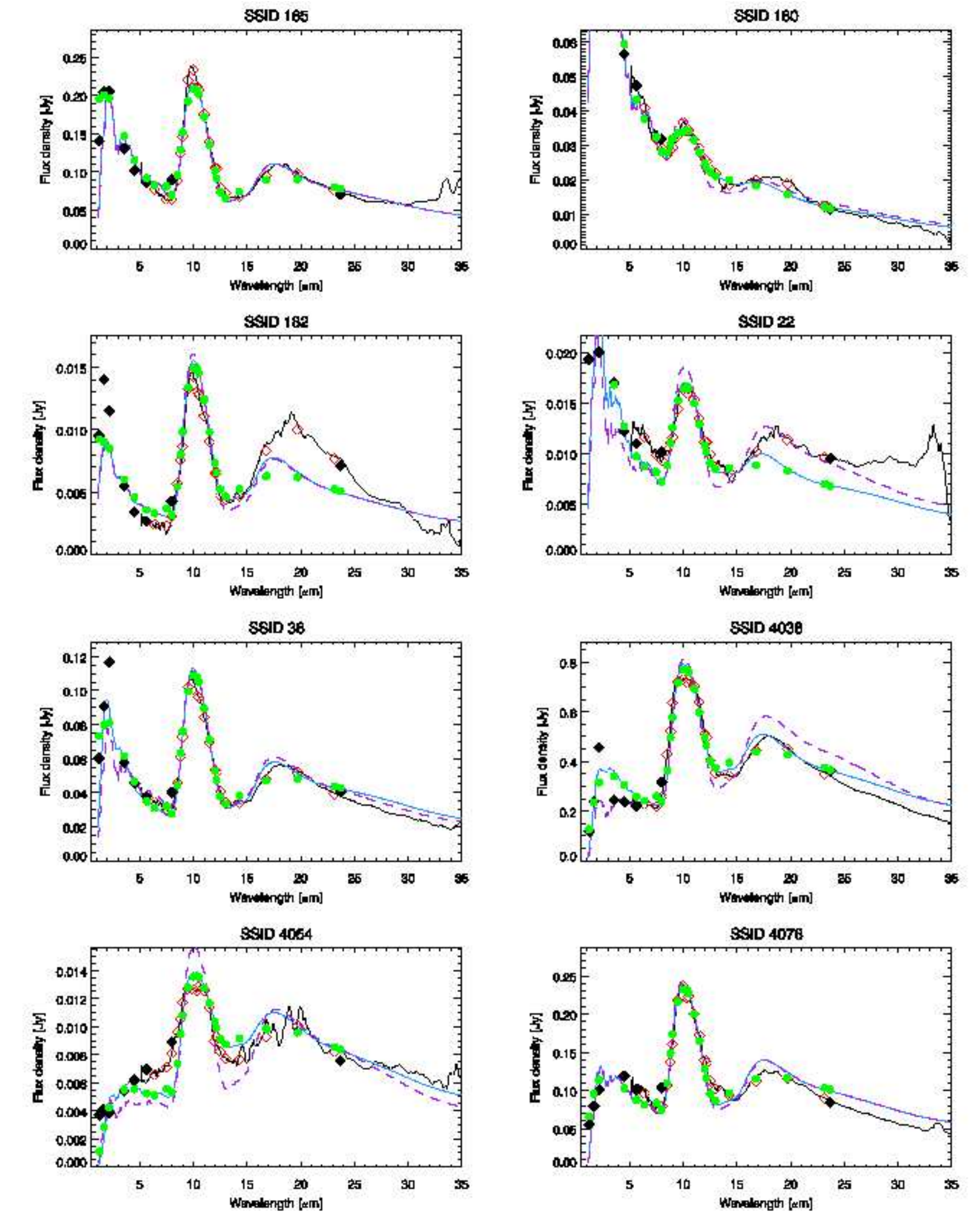}
\caption[Best-fit models to the O-AGB LMC spectra]{The observed \emph{Spitzer} spectra of O-rich AGB stars (black) and the corresponding best-fit model (blue). The photometric points are shown as black diamonds, the flux from the spectral segments as red diamonds and the synthetic flux from the models are green circles. The least-square fit is performed by matching the green model points to the red/black data points. The black and blue lines are only shown for illustration purposes. For comparison the best-fit model without an alumina component is indicated by a dashed purple line.}  
\label{fig:AlModelFit}
\end{center}
\end{figure*}

\begin{figure*}
\ContinuedFloat
\begin{center}
\includegraphics[trim=0cm 0cm 0cm 0cm, clip=true, width=\textwidth]{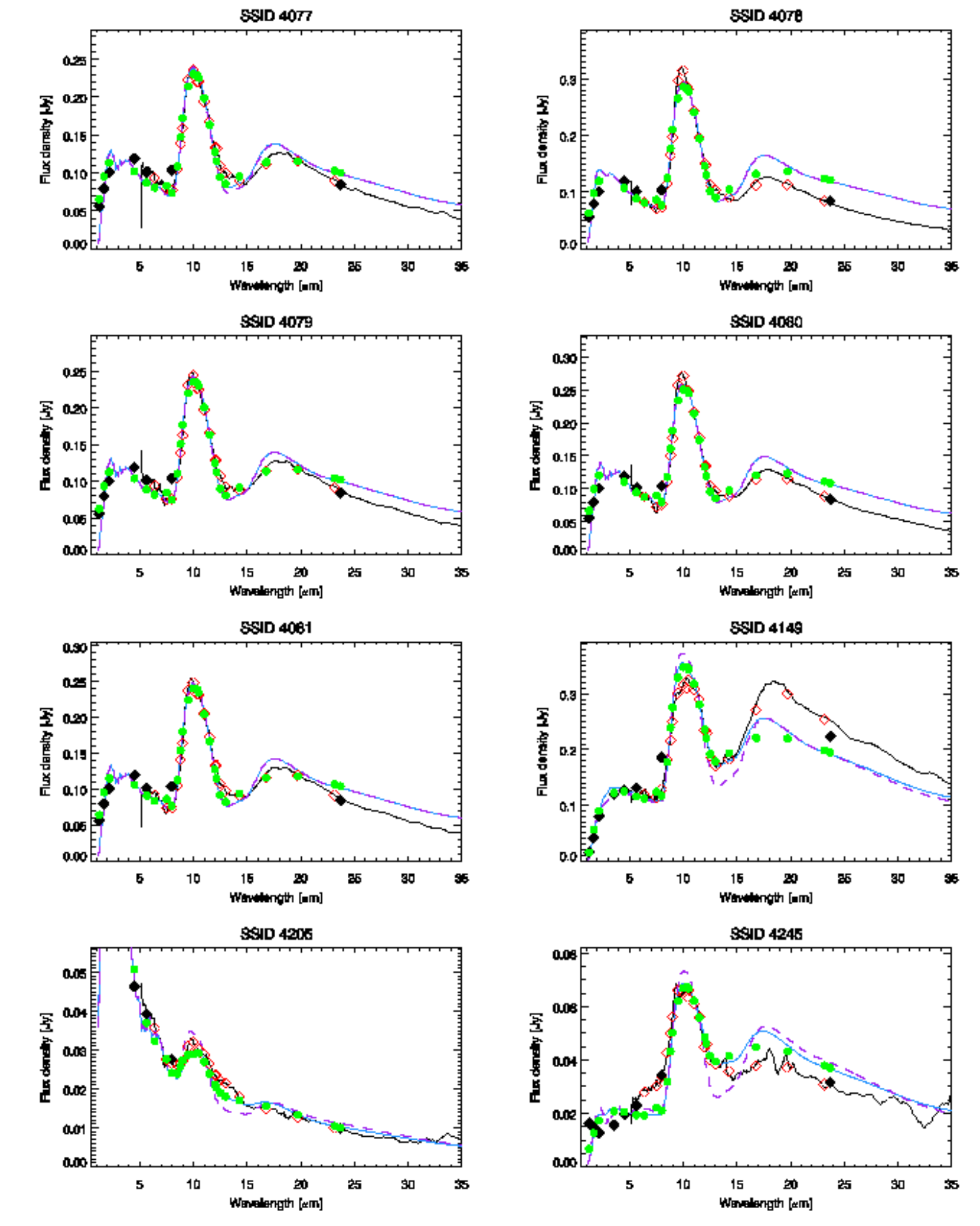}
\end{center}
\caption[]{ --- Continued.}
\end{figure*}

\begin{figure*}
\ContinuedFloat
\begin{center}
\includegraphics[trim=0cm 0cm 0cm 0cm, clip=true, width=\textwidth]{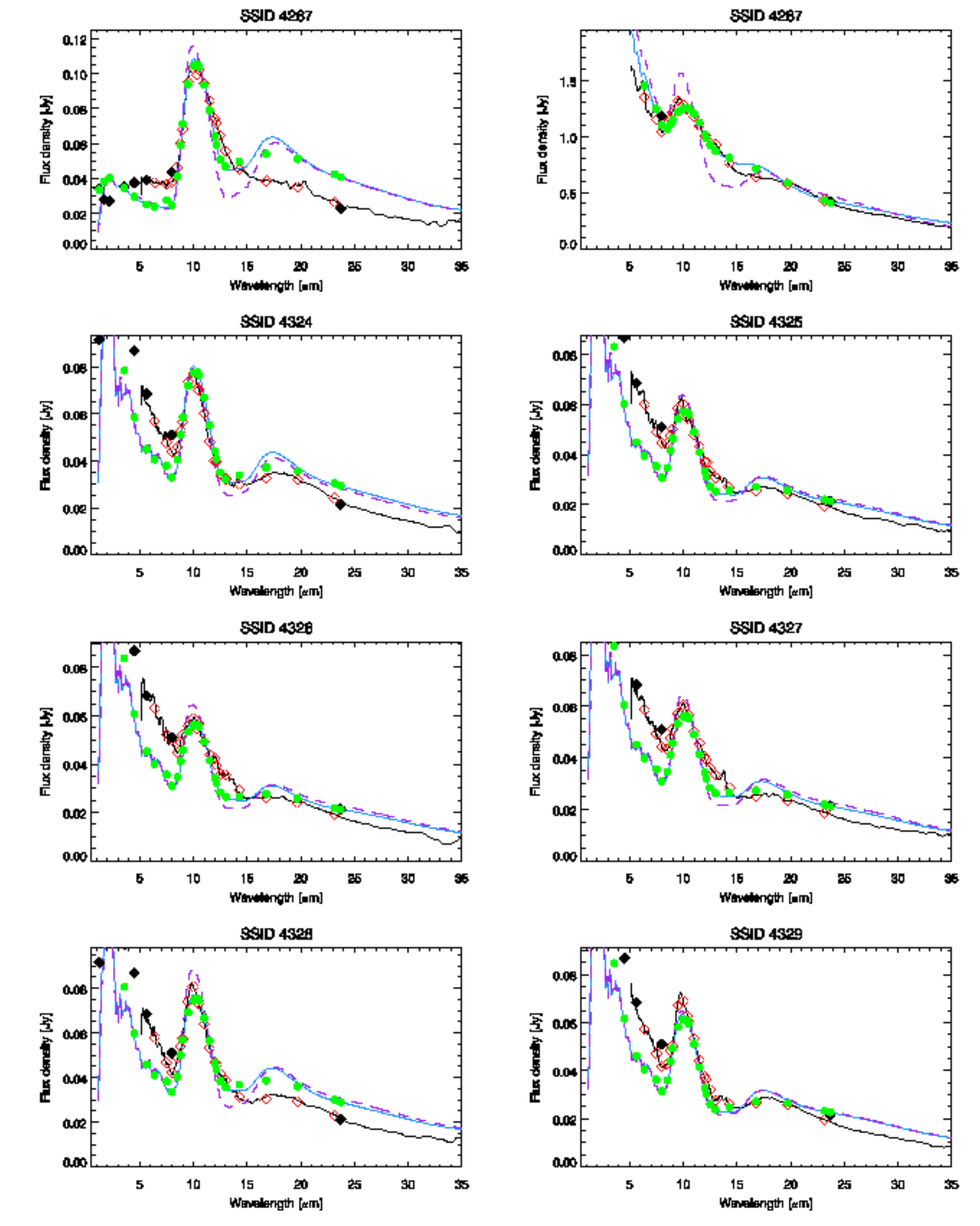}
\end{center}
\caption[]{ --- Continued.}
\end{figure*}

\begin{figure*}
\ContinuedFloat
\begin{center}
\includegraphics[trim=0cm 0cm 0cm 0cm, clip=true, width=\textwidth]{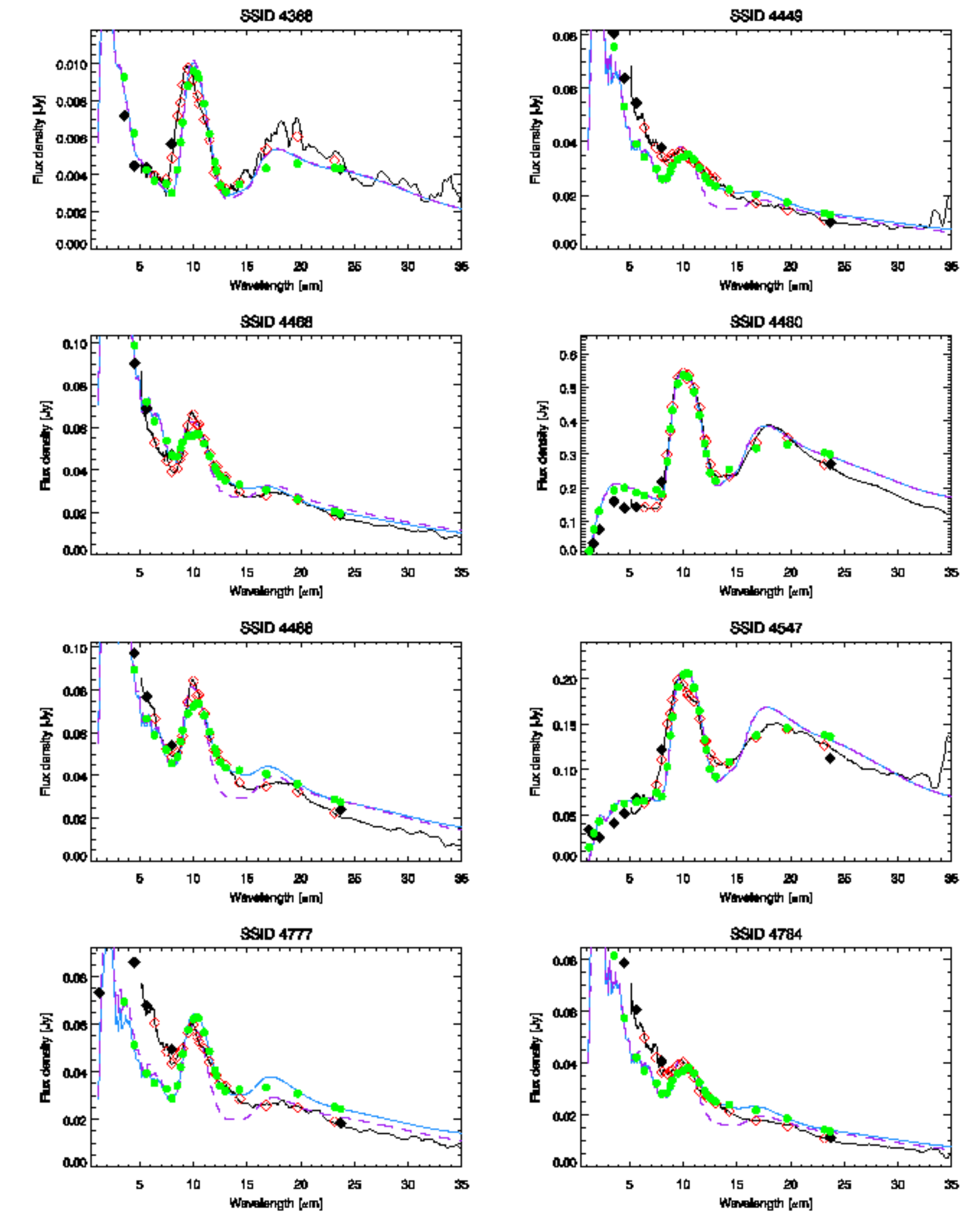}
\end{center}
\caption[]{ --- Continued.}
\end{figure*}

\begin{figure*}
\ContinuedFloat
\begin{center}
\includegraphics[trim=0cm 5cm 0cm 0cm, clip=true, width=\textwidth]{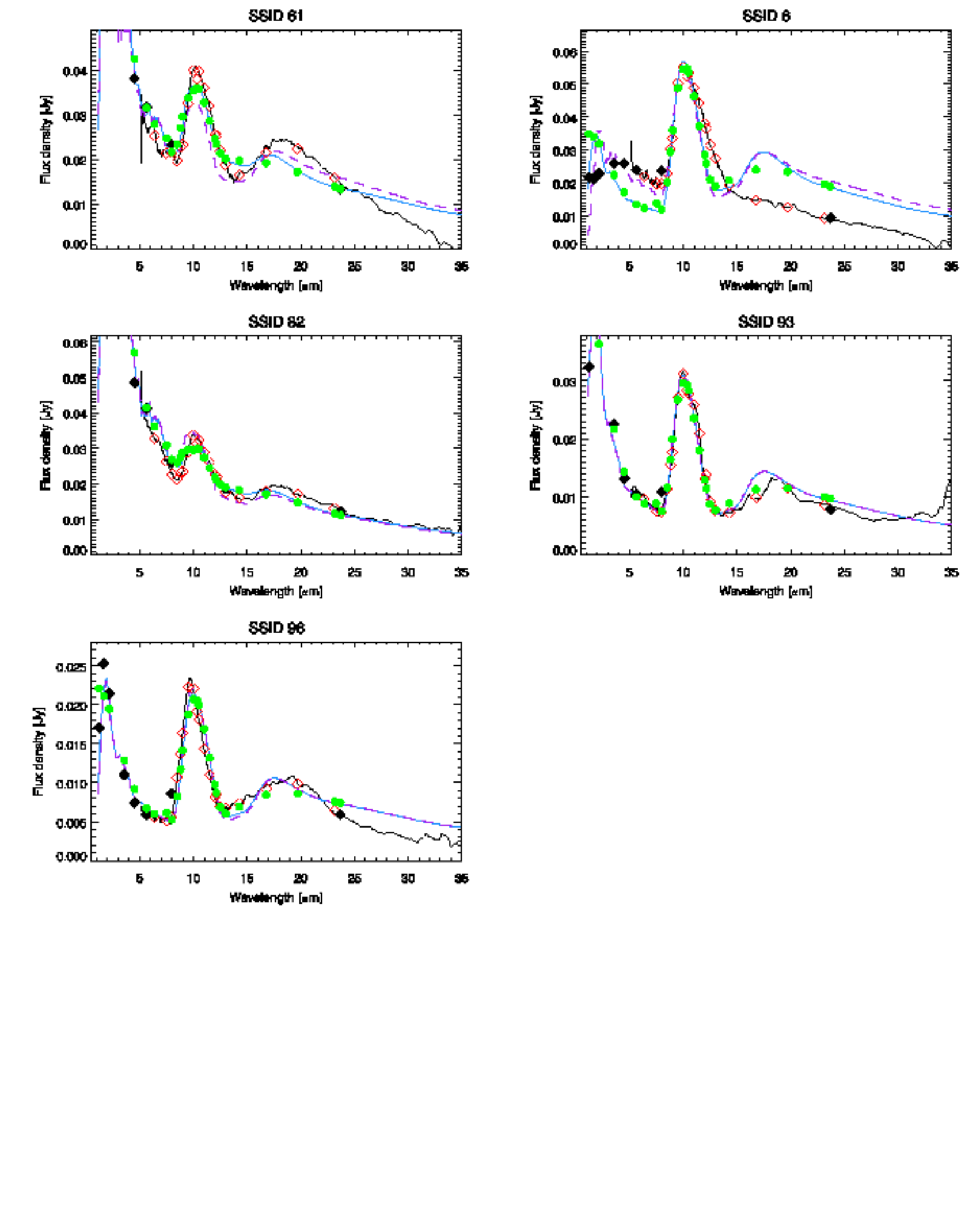}
\end{center}
\caption[]{ --- Continued.}
\end{figure*}


\begin{table*}
\centering 
 \caption{The best-fit model results for the spectra and photometry.} \label{tab:AlmodelResults}
\begin{tabular}{@{}lccccccc@{}} 
\hline\hline 
\par SSID & Spectral    & $M_{\rm bol}$ & $ R_{\rm in}$	                &	$\dot{M}$	                        &	Alx.	   &   \rchisq  &   \rchisq           \\	
	&    type	&		& (R$_{\rm star}$)		&	(M$_{\odot}\ \mathrm{yr}^{-1}$) 	&	($\%$)     &           &   No Alumina        \\	
\hline 
6	&	M1	& $-$4.69   $\pm$  0.43	&     15   $\pm$  2.5	&    (2.77 $\pm$ 0.74)$\times 10^{-7}$	&    15	  $\pm$   10	&     100.2  &   97.2      \\	
22	&	M9	& $-$4.10   $\pm$  0.26	&     2.5  $\pm$  0.5	&    (1.41 $\pm$ 0.52)$\times 10^{-7}$	&    30	  $\pm$   10	&     10.3   &   11.0      \\	
38	&	M5	& $-$5.45   $\pm$  0.36	&     5.0  $\pm$  0.5	&    (3.92 $\pm$ 0.39)$\times 10^{-7}$	&    5	  $\pm$   10	&     6.4    &   6.6       \\	
61	&	M9	& $-$5.59   $\pm$  0.19	&     3.0  $\pm$  0.5	&    (1.40 $\pm$ 0.28)$\times 10^{-7}$	&    35	  $\pm$   10	&     9.8    &   15.0      \\	
82	&	M9	& $-$5.99   $\pm$  0.22	&     7.5  $\pm$  2.5	&    (1.26 $\pm$ 0.28)$\times 10^{-7}$	&    40	  $\pm$   15	&     9.4    &   12.8      \\	
93	&	M1	& $-$4.97   $\pm$  0.07	&     15   $\pm$  2.5	&    (1.05 $\pm$ 0.11)$\times 10^{-7}$	&    0	  $\pm$   10	&     10.0   &   9.6       \\	
96	&	M5	& $-$4.03   $\pm$  0.22	&     7.5  $\pm$  2.5	&    (1.36 $\pm$ 0.37)$\times 10^{-7}$	&    5	  $\pm$   5	&     14.1   &   13.7      \\	
165	&	M9	& $-$6.43   $\pm$  0.11	&     5.0  $\pm$  2.0	&    (6.17 $\pm$ 1.13)$\times 10^{-7}$	&    0	  $\pm$   10	&     7.3    &   7.0       \\	
180	&	M9	& $-$6.01   $\pm$  0.28	&     2.5  $\pm$  2.0	&    (8.48 $\pm$ 4.08)$\times 10^{-8}$	&    40	  $\pm$   20	&     5.2    &   8.5       \\	
182     &	M1	& $-$3.24   $\pm$  0.31	&     15   $\pm$  0.5	&    (1.42 $\pm$ 0.41)$\times 10^{-7}$	&    10	  $\pm$   10	&     27.0   &   27.0      \\	
182*    &	M5	& $-$3.29   $\pm$  0.11	&     5.0  $\pm$  2.5	&    (1.45 $\pm$ 0.50)$\times 10^{-7}$	&    15	  $\pm$   15	&     11.6   &             \\	
4038	&	M5	& $-$6.96   $\pm$  0.74	&     7.5  $\pm$  2.5	&    (2.63 $\pm$ 0.52)$\times 10^{-6}$	&    20	  $\pm$   10	&     9.7    &   12.9      \\	
4054	&	M9	& $-$2.57   $\pm$  0.83	&     7.5  $\pm$  2.5	&    (6.94 $\pm$ 0.43)$\times 10^{-7}$	&    30	  $\pm$   10	&     21.8   &   22.3      \\	
4076	&	M9	& $-$5.92   $\pm$  0.10	&     5.0  $\pm$  2.0	&    (1.22 $\pm$ 0.10)$\times 10^{-6}$	&    5	  $\pm$   10	&     7.3    &   7.4       \\	
4077	&	M9	& $-$5.91   $\pm$  0.10	&     5.0  $\pm$  2.0	&    (1.22 $\pm$ 0.11)$\times 10^{-6}$	&    5	  $\pm$   10	&     6.7    &   6.7       \\	
4078	&	M5	& $-$5.91   $\pm$  0.10	&     7.5  $\pm$  2.5	&    (1.21 $\pm$ 0.10)$\times 10^{-6}$	&    0	  $\pm$   10	&     15.4   &   14.8      \\	
4079	&	M9	& $-$5.91   $\pm$  0.10	&     5.0  $\pm$  2.0	&    (1.22 $\pm$ 0.24)$\times 10^{-6}$	&    0	  $\pm$   10	&     6.4    &   6.1       \\	
4080	&	M9	& $-$5.98   $\pm$  0.13	&     5.0  $\pm$  2.5	&    (1.26 $\pm$ 0.22)$\times 10^{-6}$	&    0	  $\pm$   10	&     9.0    &   8.6       \\	
4081	&	M9	& $-$5.93   $\pm$  0.11	&     5.0  $\pm$  2.5	&    (1.23 $\pm$ 0.24)$\times 10^{-6}$	&    0	  $\pm$   10	&     7.1    &   6.8       \\	
4149	&	M1	& $-$5.76   $\pm$  0.13	&     15   $\pm$  0.5	&    (3.02 $\pm$ 0.97)$\times 10^{-6}$	&    15	  $\pm$   25	&     9.7    &   11.7      \\	
4205	&	M9	& $-$5.84   $\pm$  0.26	&     2.5  $\pm$  2.0	&    (7.85 $\pm$ 3.76)$\times 10^{-8}$	&    40	  $\pm$   20	&     6.5    &   10.7      \\	
4245	&	M5	& $-$4.02   $\pm$  0.99	&     15   $\pm$  0.5	&    (1.36 $\pm$ 0.39)$\times 10^{-6}$	&    30	  $\pm$   20	&     23.8   &   24.8      \\	
4267	&	M1	& $-$4.80   $\pm$  0.99	&     15   $\pm$  2.5	&    (7.30 $\pm$ 0.96)$\times 10^{-7}$	&    25	  $\pm$   10	&     43.9   &   48.0      \\	
4287	&	M5	& $-$10.78  $\pm$  0.38 &     5.0  $\pm$  2.5	&    (3.05 $\pm$ 1.64)$\times 10^{-7}$	&    60	  $\pm$   10	&     1.4    &   4.5       \\	
4324	&	M9	& $-$5.85   $\pm$  0.24	&     5.0  $\pm$  0.5	&    (3.15 $\pm$ 1.36)$\times 10^{-7}$	&    15	  $\pm$   10	&     18.7   &   18.7      \\	
4325	&	M9	& $-$5.92   $\pm$  0.20	&     2.5  $\pm$  2.5	&    (1.63 $\pm$ 0.41)$\times 10^{-7}$	&    15	  $\pm$   15	&     20.1   &   20.7      \\	
4326	&	M9	& $-$5.94   $\pm$  0.19	&     2.5  $\pm$  0.5	&    (1.64 $\pm$ 0.42)$\times 10^{-7}$	&    20	  $\pm$   15	&     23.9   &   25.4      \\	
4327	&	M9	& $-$5.93   $\pm$  0.20	&     2.5  $\pm$  2.0	&    (1.64 $\pm$ 0.41)$\times 10^{-7}$	&    20	  $\pm$   15	&     21.6   &   22.7      \\	
4328	&	M9	& $-$5.89   $\pm$  0.22	&     5.0  $\pm$  2.5	&    (3.21 $\pm$ 1.26)$\times 10^{-7}$	&    25	  $\pm$   10	&     20.3   &   23.0      \\	
4329	&	M9	& $-$5.94   $\pm$  0.19	&     2.5  $\pm$  2.0	&    (1.64 $\pm$ 0.42)$\times 10^{-7}$	&    5	  $\pm$   10	&     18.0   &   17.5      \\	
4368	&	M5	& $-$3.67   $\pm$  0.40	&     2.5  $\pm$  2.5	&    (5.79 $\pm$ 1.08)$\times 10^{-8}$	&    5	  $\pm$   10	&     20.2   &   19.5      \\	
4449	&	M9	& $-$5.88   $\pm$  0.26	&     3.0  $\pm$  0.5	&    (1.20 $\pm$ 0.38)$\times 10^{-7}$	&    50	  $\pm$   20	&     16.2   &   21.9      \\	
4468	&	M9	& $-$6.56   $\pm$  0.21	&     2.5  $\pm$  2.5	&    (1.09 $\pm$ 0.11)$\times 10^{-7}$	&    40	  $\pm$   15	&     9.1    &   11.9      \\	
4480	&	M5	& $-$6.11   $\pm$  0.11	&     15   $\pm$  0.5	&    (5.32 $\pm$ 1.20)$\times 10^{-6}$	&    35	  $\pm$   25	&     46.5   &   44.6      \\	
4486	&	M9	& $-$6.40   $\pm$  0.22	&     3.0  $\pm$  0.5	&    (2.03 $\pm$ 0.35)$\times 10^{-7}$	&    40	  $\pm$   15	&     6.8    &   10.4      \\	
4547	&	M9	& $-$5.24   $\pm$  0.70	&     15   $\pm$  0.5	&    (3.56 $\pm$ 1.04)$\times 10^{-6}$	&    0	  $\pm$   20	&     27.1   &   26.0      \\	
4777	&	M9	& $-$5.73   $\pm$  0.10	&     5.0  $\pm$  0.5	&    (2.98 $\pm$ 1.27)$\times 10^{-7}$	&    30	  $\pm$   15	&     33.6   &   34.5      \\	
4784	&	M9	& $-$5.96   $\pm$  0.25	&     3.0  $\pm$  0.5	&    (1.24 $\pm$ 0.40)$\times 10^{-7}$	&    50	  $\pm$   20	&     19.9   &   20.9      \\	
\hline 
\multicolumn{8}{l}{Note: The best-fit obtained without a metallic iron component is indicated by an *.}
\end{tabular} 
\end{table*}

\begin{figure*}
\begin{center}
\includegraphics[trim=0cm 0cm 0cm 0cm, clip=true, width=\textwidth]{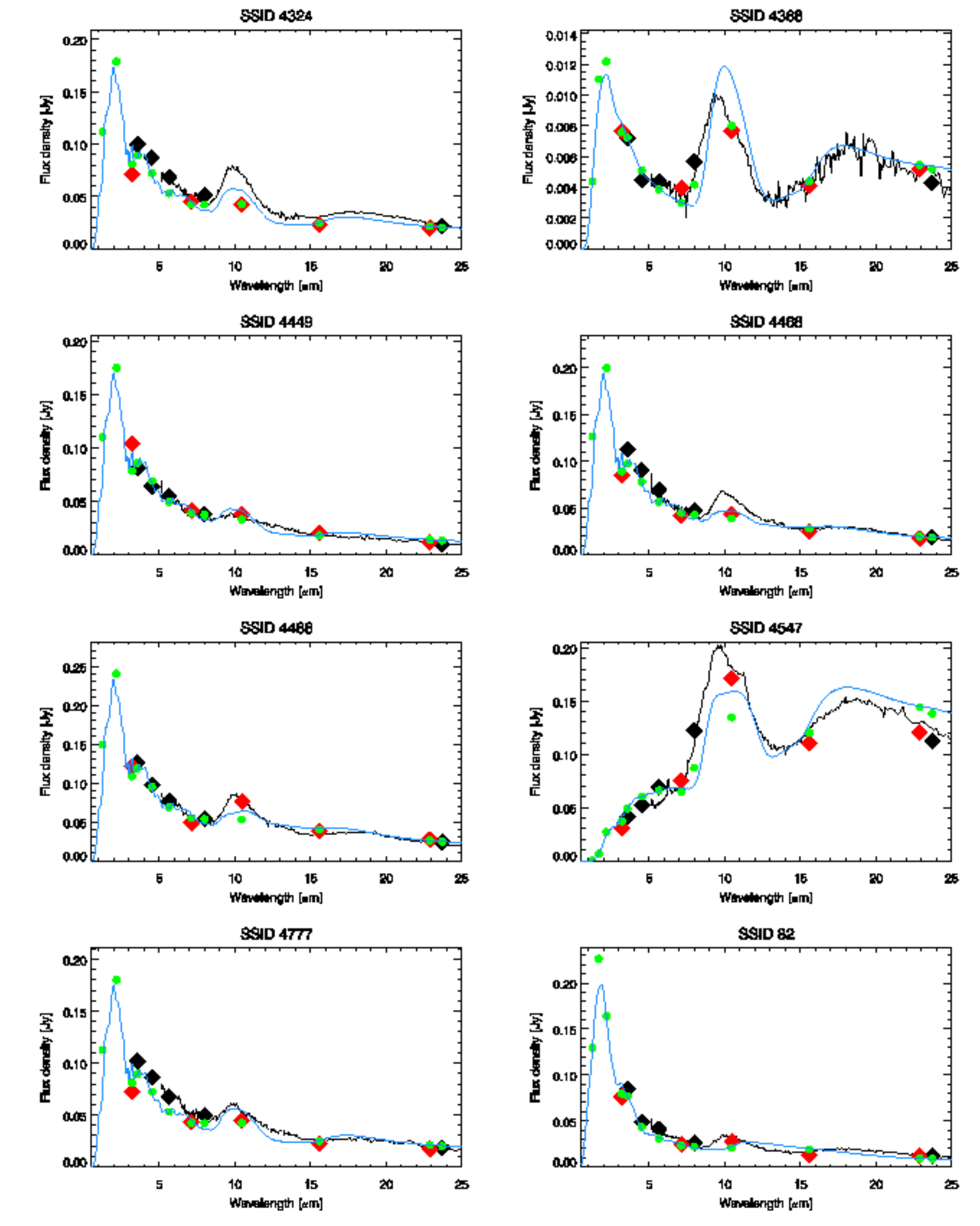}
\caption[Best-fit models to the O-AGB LMC SED]{The observed \emph{Spitzer} spectra of O-rich AGB stars (black) and the corresponding best-fit model (blue). The SAGE photometry is shown as black diamonds, the {\em AKARI} data as red diamonds and the synthetic flux from the models are green circles. The least-square fit is performed by matching the green model points to the red/black photometric data points. The spectra is shown only for illustration purposes.}
\label{fig:AlSEDModelFit}
\end{center}
\end{figure*}

 Eight stars in our spectroscopic sample have also been observed with {\em AKARI} at 3.2, 7, 11, 15 and 24 $\mu$m. This provides additional photometric coverage in the mid-infrared spectral region useful for distinguishing the dust composition. For these sources we also determine the best-fit model using only the photometric data from the {\em Spitzer} SAGE and {\em AKARI} LMC point source catalogues, to see how the model parameters (particularly the alumina abundance) are affected by the exclusion of the spectrum. The best-fit obtained using only the photometry is shown in Figure~\ref{fig:AlSEDModelFit} and the results summarised in Table~\ref{tab:AlmodelSEDResults}.

\begin{table*}
\centering 
 \caption{The best-fit SED model results.} \label{tab:AlmodelSEDResults}
\begin{tabular}{@{}lcccccc@{}} 
\hline\hline 
\par SSID & Spectral    & $M_{\rm bol}$ & $ R_{\rm in}$	                &	$\dot{M}$	                        &	Alx.	   &   \rchisq       \\	
	&    type	&		& (R$_{\rm star}$)		&	(M$_{\odot}\ \mathrm{yr}^{-1}$) 	&	($\%$)     &                  \\	
\hline 
82	&       M5   & $-$6.43  $\pm$ 0.08 &  5.0    $\pm$   2.0 	& (6.17 $\pm$ 0.95)$\times 10^{-8}$             &    100  $\pm$   25     &  21.7   \\            
4324	&       M9   & $-$6.24  $\pm$ 0.03 &  7.5    $\pm$   2.0 	& (1.89 $\pm$ 0.38)$\times 10^{-7}$	        &    0    $\pm$   20     &  33.6   \\                                 
4368	&       M1   & $-$3.10  $\pm$ 0.25 &  2.5    $\pm$   0.5 	& (8.89 $\pm$ 2.23)$\times 10^{-8}$	        &    0    $\pm$   15     &  22.4   \\ 				 
4449	&       M9   & $-$6.21  $\pm$ 0.01 &  5.0    $\pm$   0.5 	& (9.29 $\pm$ 0.16)$\times 10^{-8}$	        &    0    $\pm$   20     &  4.8    \\  				 
4468	&       M9   & $-$6.36  $\pm$ 0.02 &  7.5    $\pm$   2.5 	& (1.99 $\pm$ 0.14)$\times 10^{-7}$	        &    50   $\pm$   25     &  18.3   \\				 
4486	&       M9   & $-$6.56  $\pm$ 0.04 &  5.0    $\pm$   0.5 	& (2.18 $\pm$ 0.13)$\times 10^{-7}$	        &    60   $\pm$   20     &  14.1   \\ 				 
4547	&       M9   & $-$4.97  $\pm$ 0.14 & 15.0    $\pm$   0.5 	& (5.25 $\pm$ 0.39)$\times 10^{-6}$	        &    5    $\pm$   25     &  35.7   \\				 
4777	&       M9   & $-$6.25  $\pm$ 0.03 &  7.5    $\pm$   2.5 	& (1.89 $\pm$ 0.15)$\times 10^{-7}$	        &    5    $\pm$   25     &  28.6   \\				 
\hline 															
\end{tabular} 
\end{table*}



All the O-AGB stars in our sample exhibit significant dust emission. Although the 10-$\mu$m feature is apparent in each spectrum, it exhibits a considerable variation in shape and there is a broad variety in the other spectral features between individual stars.
%
At this stage, it is worth remembering that the model grid is intended to provide an initial estimate to the fitting parameters and that each best model fit can be optimised by fine-tuning the model values on an individual basis.  

In general the fitting of the models to the spectra plus the spectral energy distribution (SED) is of a high quality for an automated routine and the model grid is very successful at reproducing the observed shape of the 10-$\mu$m feature.  The model fits to the SED without the additional information from the spectra also reproduce the observations, however the constraints on the model parameters, in particularly the alumina abundance are less robust. We find that the combined SAGE and AKARI photometric data allows us to identify stars with a significant alumina component in the majority of sources, however for stars with a very low contrast 10-$\mu$m feature (e.g.~SSID 4449) spectroscopic information is still required. 
              
At longer wavelengths ($\lambda > 16~\mu$m) the models are less successful in reproducing the broad emission feature at 18 $\mu $m and the slope of the continuum in the observed spectra. This discrepancy is due in part to the choice of optical constants; in laboratory silicates the 18-$\mu $m emission feature peaks at too short a wavelength compared to observations and its strength is often too strong with respect to the observed 10-$\mu $m feature. Conversely, astronomical silicates provide a better fit to the 20-\mum region, however, the shape is broader than observed and they over-predict the flux in the 10-\mum region and cannot reproduce its shape and position. 
More complicated situations are not considered in the model such as: different spatial density distributions of the dust; deviations from spherical symmetry; or a steeper density profile, which may cause deviations between the model and the observations at longer wavelengths. 


In some stars (for example, SSID 4329), there is a small discrepancy between the models and the spectra in the 4--8-$\mu$m  wavelength region. The inclusion of iron grains is necessary to reproduce the general slope of the SED and explain excess emission in the spectra up to 8 $\mu $m, as laboratory silicates provide too little opacity in the near-IR \citep{Kemper2002, Verhoelst2009, McDonald2010}. Other proposed sources of this excess continuum emission includes amorphous carbon \citep{Demyk2000} or micron-sized O-rich dust grains \citep{Hoefner2008, Norris2012}. In these cases the metallic iron abundances may need to be increased in order to reproduce the observed profiles. In this study we use an iron abundance of 4 per cent with respect to the amorphous silicates, however \cite{Groenewegen2009} find a value of 5 per cent provides the best fit.

\begin{figure}
\centering
\includegraphics[trim=0cm 0cm 0cm 0cm, clip=true,width=84mm]{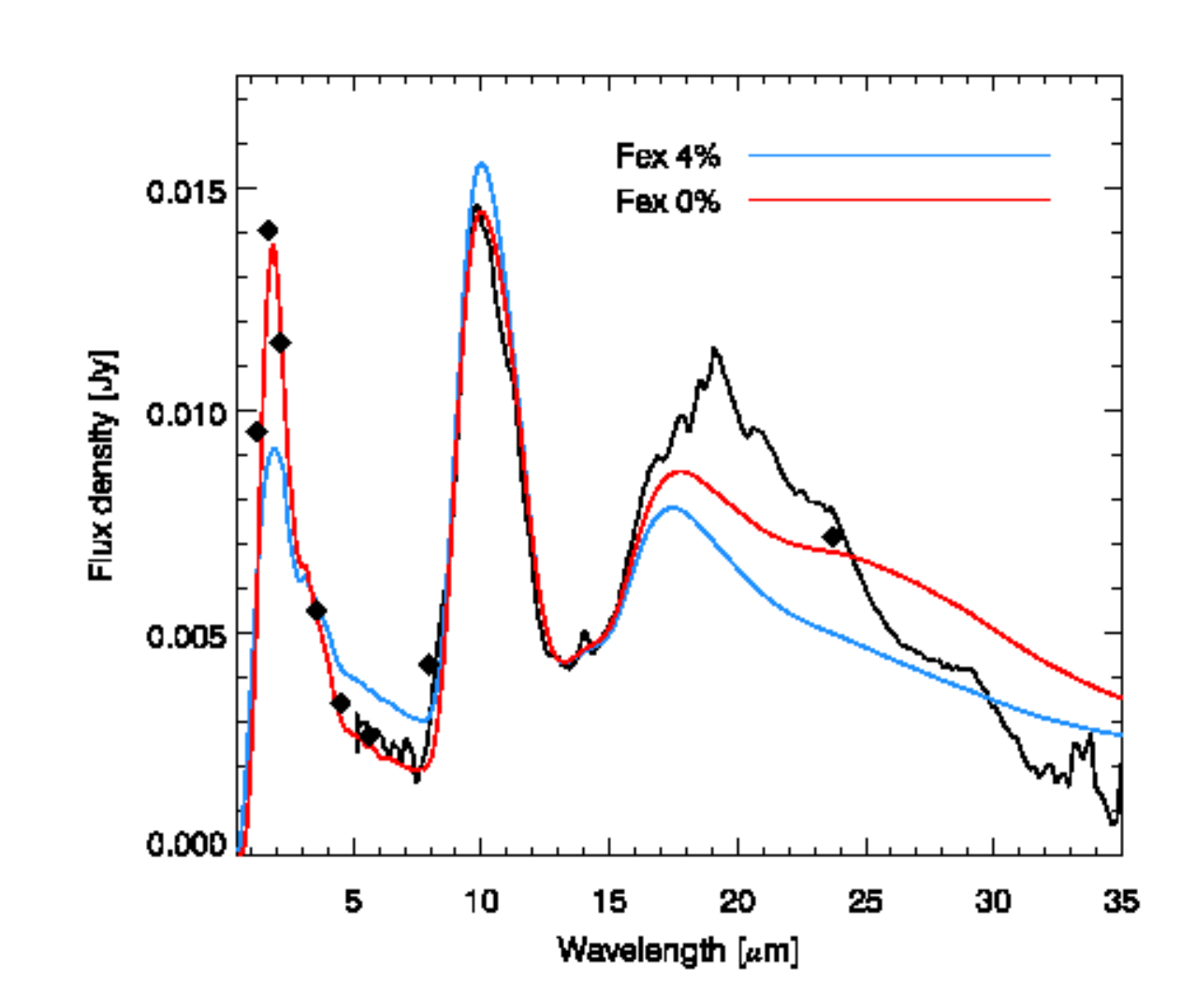}
 \caption{Fits to the SEDs and IRS spectra of SSID 182, using metallic iron (blue) and without iron (red). This is the only source where the fit is not improved with an iron component. The best-fit model obtained from our grid with 4 per cent iron grains has a \chisq of 27.0, while the best-fit without iron had a \chisq of 11.6.}
  \label{fig:SSID182_Fe_Spec} 
\end{figure}

While the adoption of iron grains has generally worked well, in one instance (SSID 182) its inclusion proves detrimental to the fit. Figure~\ref{fig:SSID182_Fe_Spec} shows the best-fitting models to SSID 182. The total mass-loss rate for both models was $\sim$1.4$\times 10^{-7}$ M$_{\odot }$ yr$^{-1}$, and the alumina mass fraction $\sim$ 15 per cent. The inclusion of metallic iron has a greater influence on the dust shell inner radius, which  determines the temperature of the hottest dust.

In our modelling we only use three dust components: amorphous silicates, amorphous alumina and metallic iron; this grain mixture fits the observed spectra well.  Other dust components that may be present but are not taken into account include crystalline silicates, gehlenite (Ca${_2}$Al${_2}$SiO${_7}$) and spinel (MgAl${_2}$O${_4}$). Three spectra (SSID 22, 182 and 419) have strong crystalline silicate complexes near 23, 28 and 33 $\mu$m, which will contribute some flux to the underlying continuum. A more detailed description of the crystalline silicate features in the present sample can be found in \cite{Jones2012}.

Previously, calcium-aluminium-rich silicates, such as gehlenite (Ca${_2}$Al${_2}$SiO${_7}$) have been detected in significant amounts (25--50 per cent) around O-rich evolved stars with low mass-loss rates \citep{Mutschke1998, Speck2000, Heras2005, Verhoelst2009}.  In these models gehlenite was included to fit the short-wavelength component of the 10-\mum feature and to provide additional flux in the 19-$\mu$m region. 
We find that the 10-\mum feature in our sample is fit reasonably well using a combination of Mg-rich silicates and alumina; thus the contribution from other dust species can only be small. This discrepancy may be due to the choice of optical constants use for the amorphous silicates.


\begin{figure}
\centering
\includegraphics[trim=0.5cm 0cm 0cm 0cm, clip=true,width=84mm]{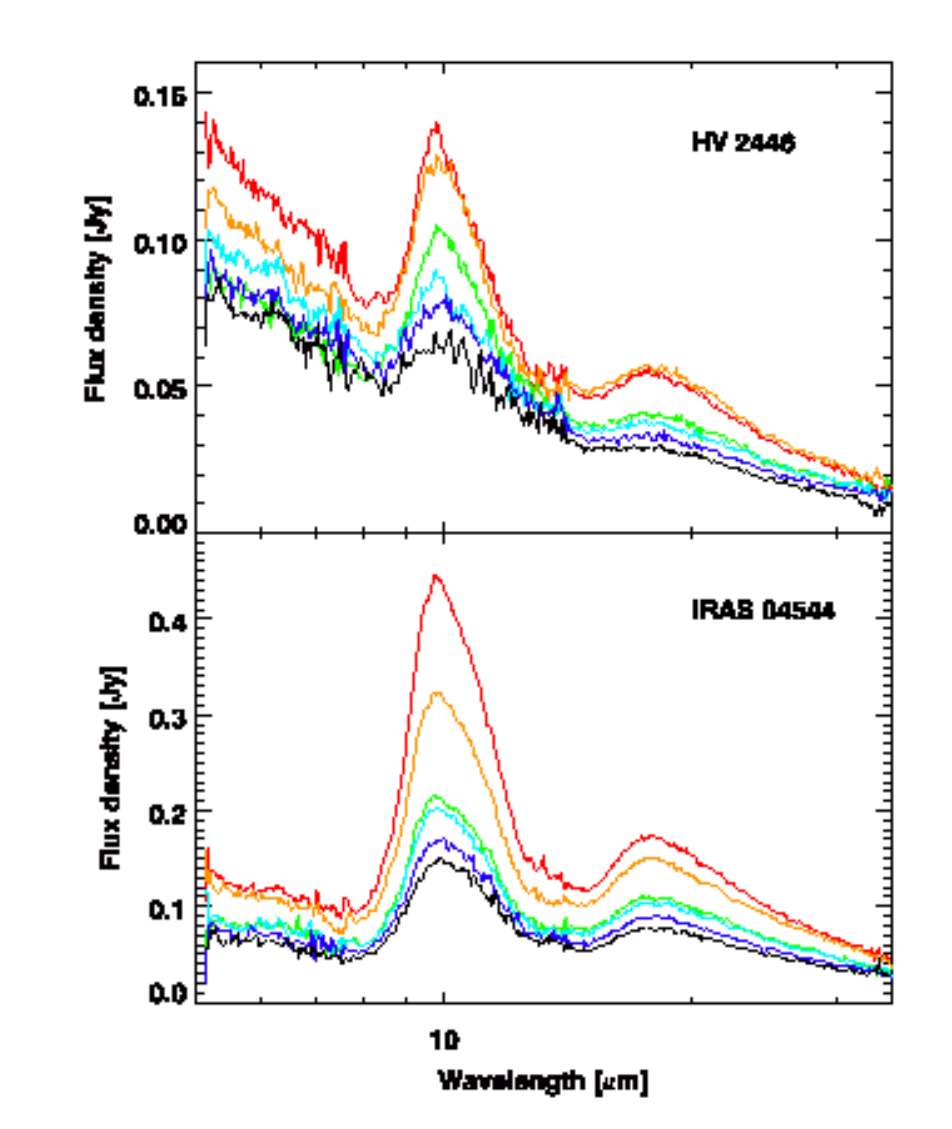}
 \caption[Observed spectral variations with pulsation phase]{Observed spectral variations for HV 2446 (SSID 4324--4329) and IRAS 04544$-$6849 (SSID 4076--4081).}
  \label{fig:specVar}
\end{figure}

Two sources in our sample, IRAS 04544$-$6849 (SSID 4076--4081) and  HV 2446 (SSID 4324--4329), have been observed with the IRS at different epochs over one variability cycle. The SED for each source includes photometry from multiple epochs, while the spectra represent an instantaneous view of the star at a precise phase of the pulsation cycle. In both these stars the strength ratio of the 10- to 20-$\mu$m silicate emission features varies with pulsation phase (Figure~\ref{fig:specVar}).  This is due to changes in the dust temperature, which is directly related to the luminosity of the star \citep{Monnier1998}. 
For both these sources the stellar and dust-shell parameters of our best-fit models are consistent across the pulsation cycle, as to some extent temporal variations are minimised. However, our models are sensitive to the change in dust optical properties, as the peak wavelength and width of the 10-$\mu$m feature in the spectra shows discernible variations between maximum and minimum light in the pulsation cycle; this indicates that the dust properties in the shell vary with pulsation phase. Furthermore, it will introduce some scattering in our model results when comparing between stars.

The brightest star in the sample, RS Men (SSID 4287), has a bolometric magnitude in excess of $-$10.78 if it was at the distance of the LMC, which is significantly above the luminosity expected for a bright RSG. This star has been identified as a foreground Mira variable with a distance of 4.75 kpc and radial velocity of 140 km s$^{-1}$ \citep{Buchanan2006, Whitelock1994}. Incidentally, this source also has the highest alumina fraction in our sample of 60 per cent.


SSID 6 along with SSID 4267 are not well described by our models and their fits are the least satisfactory. These spectra are unusual as they exhibit a strong 10-\mum silicate feature but no 18-\mum feature. One possible explanation is that the silicates have recently formed and the  18-\mum feature has a low contrast ratio with respect to the 10-\mum feature. Alternatively this may be explained by a lack of cooler dust, or the dust may be composed only of small 0.1 \mum Mg-rich olivine and alumina grains \citep{Gielen2011}.

\section{Discussion}\label{Discussion}

\begin{figure}
\centering
\includegraphics[trim=0cm 0cm 0cm 0cm, clip=true,width=84mm]{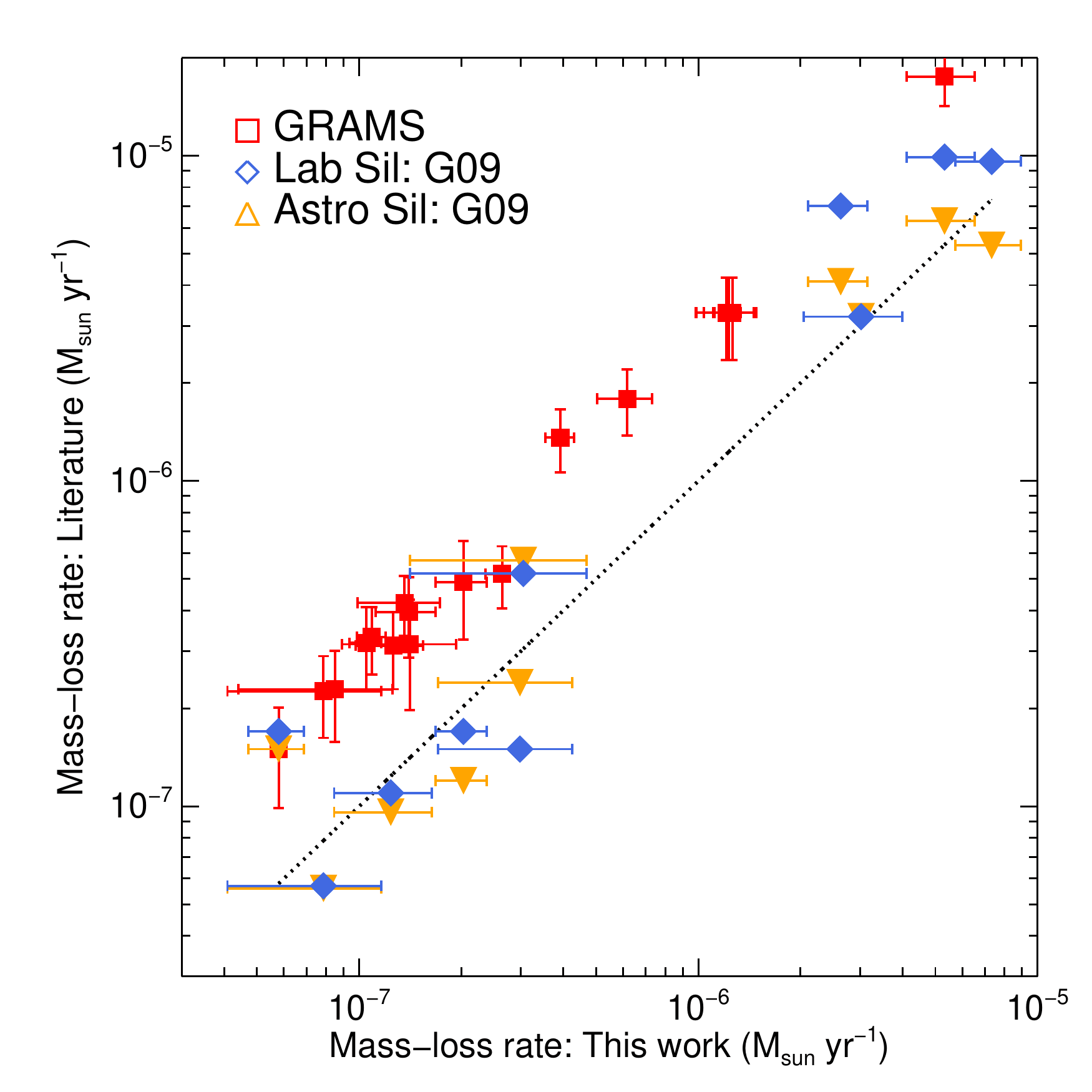}
 \caption[Comparison of model MLRs with literature values]{Comparison of mass-loss rated derived in this work with objects in common with our sample from \cite{Groenewegen2009} and the {\sc grams} model results from \cite{Riebel2012}.}
  \label{fig:litMLRcomp}
\end{figure}

Using the dust-shell models presented here, we are able to constrain both the mass-loss rate and the fractional abundance of alumina. Based on the best-fitting models the mass-loss rates range from $\sim 8\times10^{-8}$ to $5\times10^{-6}$ M$_{\odot}\ \mathrm{yr}^{-1}$, and the alumina fraction ranges from 0 to 50 per cent.  In Figure~\ref{fig:litMLRcomp} we compare our derived mass-loss rates to those determined from SED fitting with the {\sc grams} model in \cite{Riebel2012} and to values listed in \cite{Groenewegen2009}. The agreement is generally quite good between our mass-loss rates and the literature values, however, our mass-loss rates are slightly lower than the {\sc grams} fits, predominantly due to the inclusion of metallic iron grains and alumina in our models.

To have a better understanding of the mass-loss rates and dust abundances derived by this fitting, some limitations of the models must be kept in mind.
The photospheric models used are appropriate for solar metallicities, and may not be entirely representative of the molecular abundances typical of AGB stars in the Magellanic Clouds. This is more significant for stars with a very low contrast dust excess (MLR $\lesssim 5\times10^{-8}$ M$_{\odot}\ \mathrm{yr}^{-1}$), where there is a strong photospheric contribution and molecular absorption features due to CO, H${_2}$O, and SiO are prominent. 
Observing the CO emission lines in the submillimetre (due to rotational transitions) for a number of Magellanic Cloud sources would also be beneficial to constrain the mass-loss parameters. This would enable us to have a better understanding of the expansion velocity and gas-to-dust ratios in the envelopes of evolved stars. 
The fits to the sources may also be improved by including a finer grid spacing, for instance, the iron abundance could be incremented in one percent intervals (up to abundances of 10 per cent) to improve the fit to the stars in the near-IR region.


In our sample, the dust composition is best fit by amorphous silicates, with appreciable amounts of amorphous alumina and additional, small contributions from metallic iron. Alumina is detected in most sources, however, in all cases its fractional abundance is less than 50 per cent. 
The alumina fractions we have obtained are consistent with those estimated using other methods (see Sections~\ref{sec:ALgrid_CCDs} and \ref{sec:SEidx}).

\begin{figure}
\centering
\includegraphics[trim=1cm 0cm 0cm 0cm, clip=true,width=84mm]{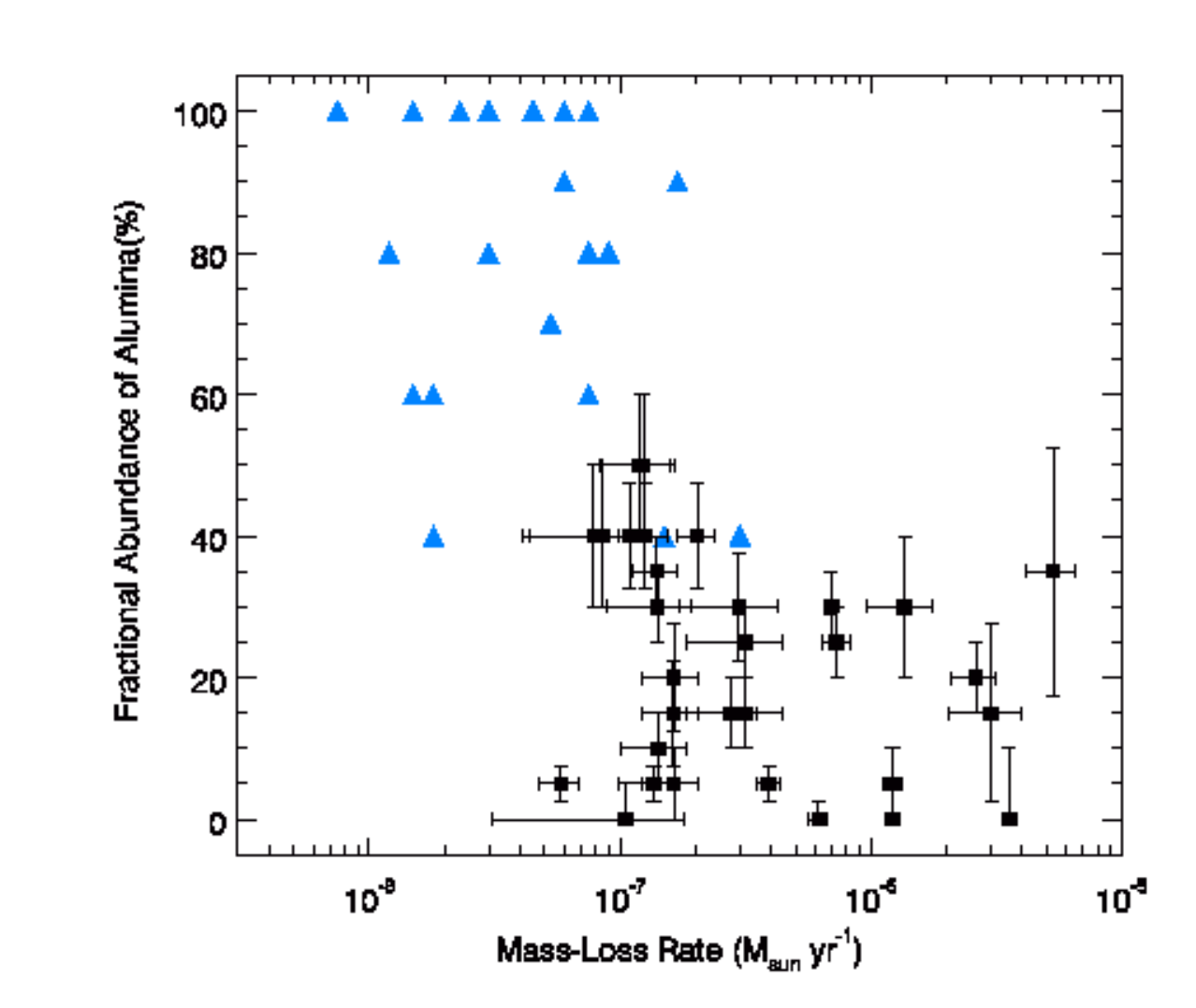}
 \caption[The fractional abundance of alumina in the dust shell as a function of mass-loss rate]{The fractional abundance of alumina in the dust shell of LMC O-AGB stars as a function of mass-loss rate (black squares). Based on Spearman's $\rho$ statistics, the correlation coefficient for the LMC sources is $-0.41$ with a significance value of 0.01. For comparison the blue triangles show the distribution of Galactic Bulge O-AGB sources from \cite{Blommaert2006}.}
  \label{fig:AlxMLR}
\end{figure}

Figure~\ref{fig:AlxMLR} shows the fractional abundance of alumina plotted against the mass-loss rate. As noted by other authors \citep{Heras2005, Blommaert2006, Lebzelter2006}, there is a correlation between the dust composition and mass-loss rate; the lower the mass-loss rate the higher the percentage of alumina in the shell.



Stars in the early stages of the AGB phase, tend to be less luminous and have lower mass-loss rates than their more evolved counterparts. This makes reliable determination of the dust composition challenging, especially for spectra with a low signal-to-noise ratio or with a low contrast dust excess. 
AGB stars dominated by alumina dust tend to have mass-loss rates below $10^{-7} \, \mathrm{M}_{\odot}\ \mathrm{yr}^{-1}$ \citep{Blommaert2006}. Indeed, the SEDs of a large fraction of Galactic Bulge sources in this mass-loss range can be fit solely with amorphous alumina. 
The oxygen-rich AGB stars in our LMC sample, which show a dust emission contrast, tend to be more luminous, with $M_{\rm init}$ of $\sim$5 $M_\odot $. 
 Our sample is dominated by stars with mass-loss rates between  $10^{-7}$ and 5 $\times 10^{-6}\,  \mathrm{M}_{\odot}\ \mathrm{yr}^{-1}$. These stars display dust components which are from the more advanced stages of the dust condensation sequence; indicated by the low percentage of alumina dust. 
It should therefore come as no surprise that silicates are the dominant dust type in the present sample. Metallicity effects on the dust-formation efficiency in oxygen-rich stars may also be important. 

Future telescopes such as the {\em JWST} and {\em SPICA} will be able to obtain high-quality spectra of low mass-loss rate, low luminosity AGB stars in the Magellanic Clouds and other resolved stellar populations, eliminating the observational bias towards silicate-rich stars in the current sample. We expect these stars to contain a high percentage of alumina-rich dust.

\section{Conclusions}\label{conclusions}

We have presented a grid of dust radiative transfer models which explores a range of alumina and silicate dust compositions and stellar and dust shell parameters. The models have been used to simultaneously fit the spectra and broadband SED of  37 oxygen-rich AGB stars in the LMC with optically-thin circumstellar envelopes.
The mass-loss rates of our sample range from $\sim 8\times10^{-8}$ to $5\times10^{-6}$ M$_{\odot}\ \mathrm{yr}^{-1}$. We find that a combination of amorphous silicates, amorphous alumina and metallic iron provides a good fit to the spectra. All our sources we found to be silicate-rich, though alumina and iron is often present in significant amounts. This dust composition is consistent with the thermodynamic dust condensation sequence for oxygen-rich AGB stars.
Furthermore, we show from dust models that the {\em AKARI} [11]--[15] versus [3.2]--[7] colour, is able to determine the fractional abundance of alumina in oxygen-rich AGB stars.

\section*{Acknowledgements}
We would like to thank Ben Sargent for his advice on the silicate dust optical constants. 
OJ acknowledges the support of an STFC studentship and thanks ASIAA for their support and hospitality during the completion of part of this work.
FK acknowledges support from the National Science Council under grant number NSC100-2112-M-001-023-MY3. 
This work is based on observations made with the {\em Spitzer} Space Telescope, which is operated by the Jet Propulsion Laboratory, California Institute of Technology under NASA contract 1407; and has made use of the SIMBAD and VizieR databases, CDS, Strasbourg, France.

\appendix
\section{Supporting information}

\begin{figure*}
\centering
\includegraphics[width=84mm]{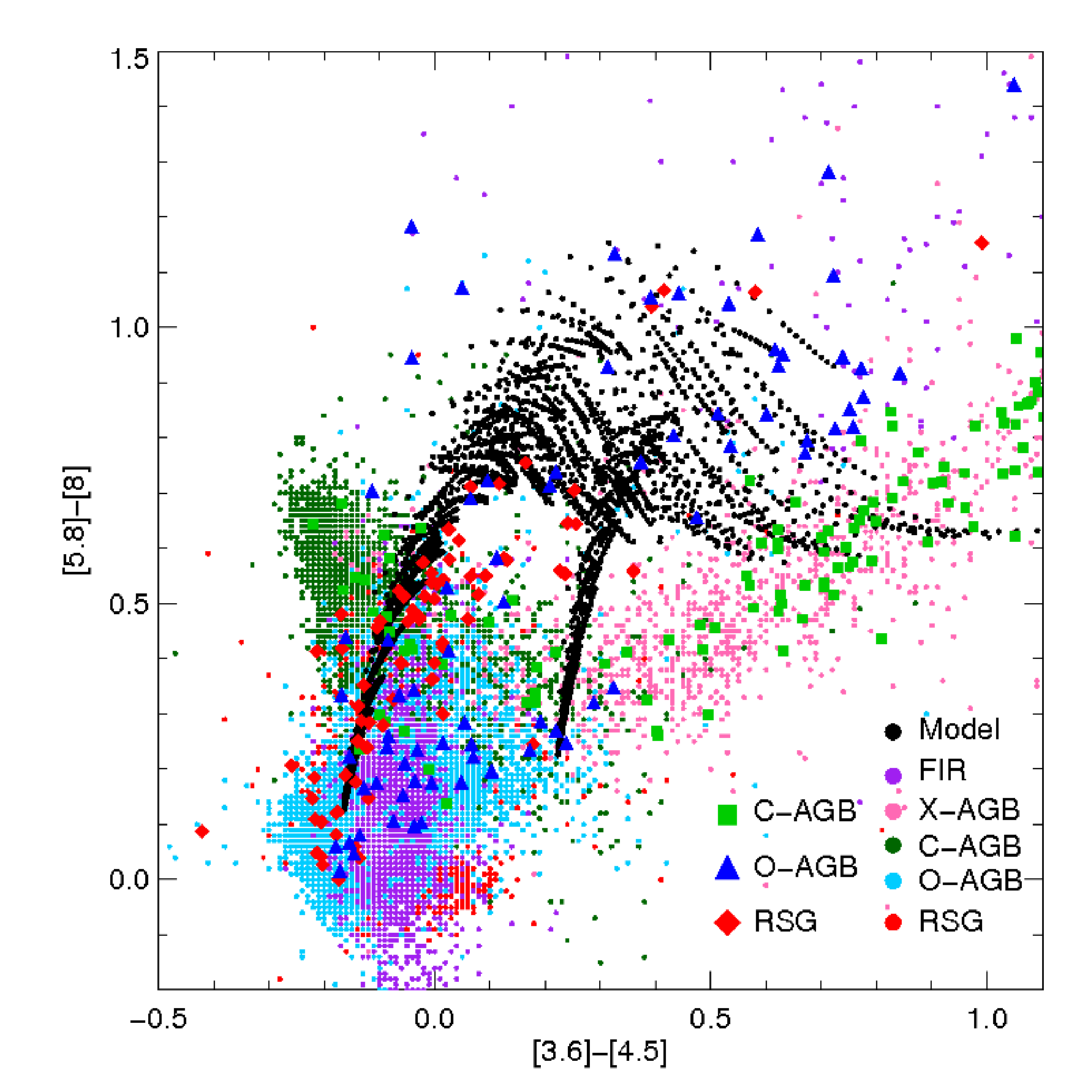}
\includegraphics[width=84mm]{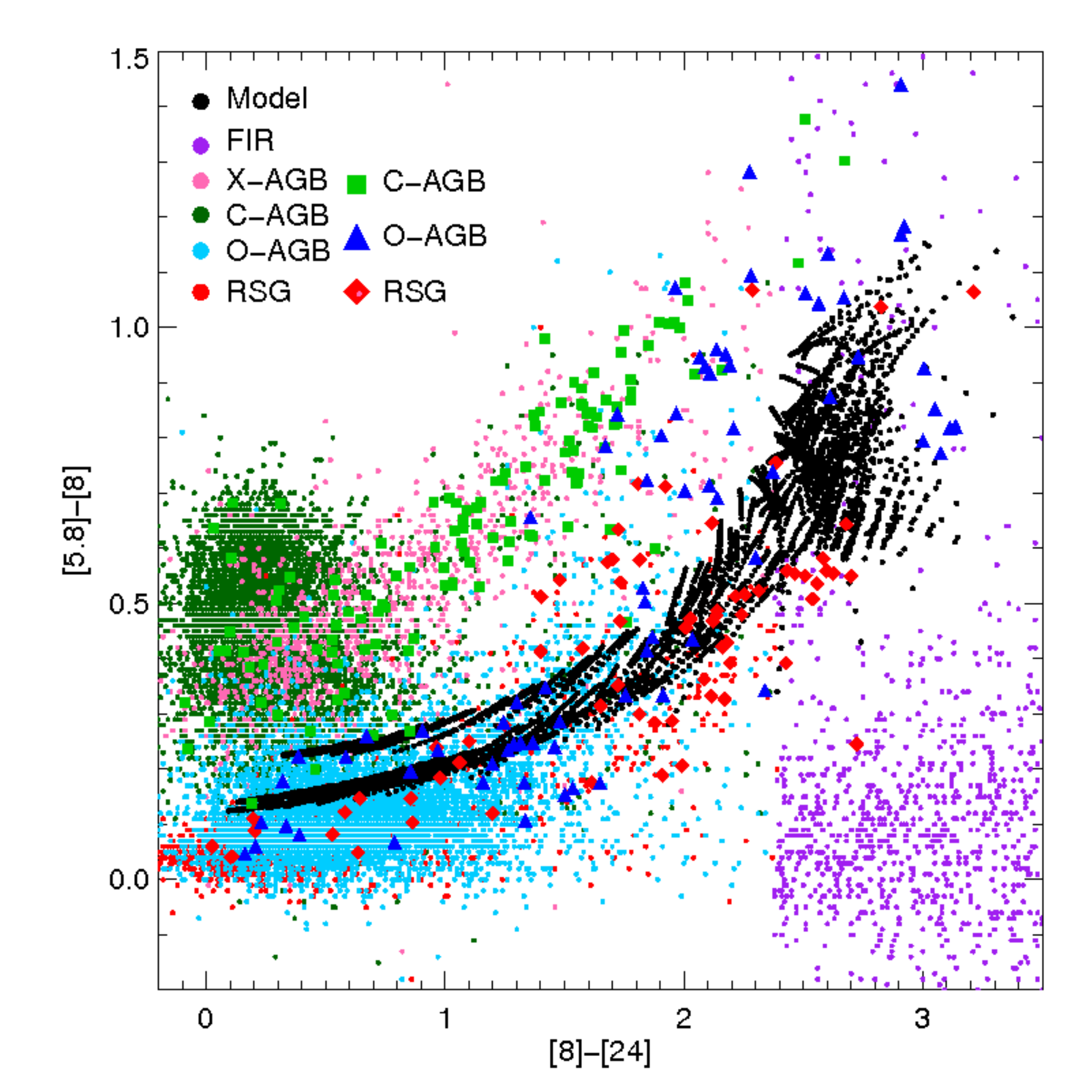}
\caption[SAGE colour-colour diagrams illustrating model grid coverage]{{\em Spitzer} IRAC/MIPS colour-colour diagrams showing the evolved stellar population of the LMC and their various photometric classifications, compared to our model grid. The colour of the small points represents the photometric classification of sources in the SAGE database and the small black dots denote the models from our O-rich grid. The spectroscopically classified evolved stars symbols are as in Figure \ref{fig:Al_grid_CCDs}.}
\label{fig:CCD_allEvolSAGE}
\end{figure*}

In Figure~\ref{fig:CCD_allEvolSAGE}, we show colour-colour diagrams of the model grid compared to all the evolved star candidates in the SAGE LMC sample and the confirmed O-rich and C-rich AGB stars in the SAGE-Spec sample. The O-rich AGB candidates, RSG candidates identified by \cite{Blum2006} are predominantly on the early-AGB and have a limited circumstellar excess. Photometrically the reddest sources (for which our models are optimised) are not included in the O-AGB classification, instead these objects are classified as extreme AGB stars, far-IR objects (which also includes YSOs) or are missed by the colour cuts. For the faint O-rich AGB stars the three-sigma error in the colours is sufficient to explain the scatter.
 Table \ref{tab:LMCEvolPhotometry} provides the average photometry in each band and the spectral classification of all the O-rich sources used in this study.

\begin{table*}
\centering 
 \caption{Spectral classification and photometric catalogue of all the O-rich sources used in this study. Magnitudes in this table have not been corrected for reddening. A full version is available in the electronic edition.} \label{tab:LMCEvolPhotometry}
\begin{tabular}{@{}c@{\ }c@{\ }c@{\ }c@{\ }c@{\ }c@{\ }c@{\ }c@{\ }c@{\ }c@{\ }c@{\ }c@{\ }c@{\ }c@{\ }c@{\ }c@{\ }c@{\ }c@{\ }c@{}} 
\hline\hline 
\par SSID	&	Spectral	&	2MASS	&	 	&		&	IRAC	&		&		&		&	MIPS	&	WISE	&		&		&		&	AKARI	&		&		&		&		\\
	& Class		&	J	&	H 	&	K	&	[3.6]	&	[4.5]	&	[5.8]	&	[8.0]	&	[24]	&	W1	&	W2	&	W3	&	W4	&	N3	&	S7	&	S11	&	L15	&	L24	\\
\hline
1	&	O-AGB	&	12.41	&	11.49	&	11.19	&	10.91	&	10.81	&	10.69	&	10.49	&	9.64	&	10.96	&	10.89	&	10.24	&	9.35	&	\ldots	&	\ldots	&	\ldots	&	\ldots	&	\ldots	\\
4	&	RSG	&	10.10	&	9.27	&	9.01	&	8.87	&	9.04	&	8.87	&	8.75	&	7.55	&	8.85	&	9.04	&	8.46	&	7.79	&	\ldots	&	\ldots	&	\ldots	&	\ldots	&	\ldots	\\
6	&	O-AGB	&	12.17	&	11.71	&	11.16	&	10.08	&	9.60	&	9.22	&	8.56	&	7.20	&	10.36	&	9.71	&	7.51	&	6.82	&	\ldots	&	\ldots	&	\ldots	&	\ldots	&	\ldots	\\
8	&	O-AGB	&	14.36	&	13.15	&	12.48	&	11.13	&	10.61	&	10.03	&	9.19	&	7.22	&	11.20	&	10.52	&	8.42	&	7.31	&	\ldots	&	\ldots	&	\ldots	&	\ldots	&	\ldots	\\
\hline
\end{tabular} 
\end{table*}



\def\aj{AJ}					
\def\actaa{Acta Astron.}                        
\def\araa{ARA\&A}				
\def\apj{ApJ}					
\def\apjl{ApJL}					
\def\apjs{ApJS}					
\def\ao{Appl.~Opt.}				
\def\apss{Ap\&SS}				
\def\aap{A\&A}					
\def\aapr{A\&A~Rev.}				
\def\aaps{A\&AS}				
\def\azh{AZh}					
\def\baas{BAAS}					
\def\jrasc{JRASC}				
\def\memras{MmRAS}				
\def\mnras{MNRAS}				
\def\pra{Phys.~Rev.~A}				
\def\prb{Phys.~Rev.~B}				
\def\prc{Phys.~Rev.~C}				
\def\prd{Phys.~Rev.~D}				
\def\pre{Phys.~Rev.~E}				
\def\prl{Phys.~Rev.~Lett.}			
\def\pasp{PASP}					
\def\pasj{PASJ}					
\def\qjras{QJRAS}				
\def\skytel{S\&T}				
\def\solphys{Sol.~Phys.}			
\def\sovast{Soviet~Ast.}			
\def\ssr{Space~Sci.~Rev.}			
\def\zap{ZAp}					
\def\nat{Nature}				
\def\iaucirc{IAU~Circ.}				
\def\aplett{Astrophys.~Lett.}			
\def\apspr{Astrophys.~Space~Phys.~Res.}		
\def\bain{Bull.~Astron.~Inst.~Netherlands}	
\def\fcp{Fund.~Cosmic~Phys.}			
\def\gca{Geochim.~Cosmochim.~Acta}		
\def\grl{Geophys.~Res.~Lett.}			
\def\jcp{J.~Chem.~Phys.}			
\def\jgr{J.~Geophys.~Res.}			
\def\jqsrt{J.~Quant.~Spec.~Radiat.~Transf.}	
\def\memsai{Mem.~Soc.~Astron.~Italiana}		
\def\nphysa{Nucl.~Phys.~A}			
\def\physrep{Phys.~Rep.}			
\def\physscr{Phys.~Scr}				
\def\planss{Planet.~Space~Sci.}			
\def\procspie{Proc.~SPIE}			
\def\icarus{Icarus}
\let\astap=\aap
\let\apjlett=\apjl
\let\apjsupp=\apjs
\let\applopt=\ao


\bibliographystyle{aa}
\bibliography{libby}

\begin{thebibliography}{83}
\expandafter\ifx\csname natexlab\endcsname\relax\def\natexlab#1{#1}\fi

\bibitem[{{Begemann} {et~al.}(1997){Begemann}, {Dorschner}, {Henning},
  {Mutschke}, {Guertler}, {Koempe}, \& {Nass}}]{Begemann1997}
{Begemann}, B., {Dorschner}, J., {Henning}, T., {et~al.} 1997, \apj, 476, 199

\bibitem[{{Bloecker}(1995)}]{Bloecker1995}
{Bloecker}, T. 1995, \aap, 297, 727

\bibitem[{{Blommaert} {et~al.}(2006){Blommaert}, {Groenewegen}, {Okumura},
  {Ganesh}, {Omont}, {Cami}, {Glass}, {Habing}, {Schultheis}, {Simon}, \& {van
  Loon}}]{Blommaert2006}
{Blommaert}, J.~A.~D.~L., {Groenewegen}, M.~A.~T., {Okumura}, K., {et~al.}
  2006, \aap, 460, 555

\bibitem[{{Blum} {et~al.}(2006){Blum}, {Mould}, {Olsen}, {Frogel}, {Werner},
  {Meixner}, {Markwick-Kemper}, {Indebetouw}, {Whitney}, {Meade}, {Babler},
  {Churchwell}, {Gordon}, {Engelbracht}, {For}, {Misselt}, {Vijh}, {Leitherer},
  {Volk}, {Points}, {Reach}, {Hora}, {Bernard}, {Boulanger}, {Bracker},
  {Cohen}, {Fukui}, {Gallagher}, {Gorjian}, {Harris}, {Kelly}, {Kawamura},
  {Latter}, {Madden}, {Mizuno}, {Mizuno}, {Nota}, {Oey}, {Onishi}, {Paladini},
  {Panagia}, {Perez-Gonzalez}, {Shibai}, {Sato}, {Smith}, {Staveley-Smith},
  {Tielens}, {Ueta}, {Van Dyk}, \& {Zaritsky}}]{Blum2006}
{Blum}, R.~D., {Mould}, J.~R., {Olsen}, K.~A., {et~al.} 2006, \aj, 132, 2034

\bibitem[{{Bohren} \& {Huffman}(1983)}]{BohrenHuffman1983}
{Bohren}, C.~F. \& {Huffman}, D.~R. 1983, {Absorption and scattering of light
  by small particles}, ed. {Bohren, C.~F.~\& Huffman, D.~R.}

\bibitem[{{Bouwman} {et~al.}(2000){Bouwman}, {de Koter}, {van den Ancker}, \&
  {Waters}}]{Bouwman2000}
{Bouwman}, J., {de Koter}, A., {van den Ancker}, M.~E., \& {Waters},
  L.~B.~F.~M. 2000, \aap, 360, 213

\bibitem[{{Bouwman} {et~al.}(2001){Bouwman}, {Meeus}, {de Koter}, {Hony},
  {Dominik}, \& {Waters}}]{Bouwman2001}
{Bouwman}, J., {Meeus}, G., {de Koter}, A., {et~al.} 2001, \aap, 375, 950

\bibitem[{{Bowen} \& {Willson}(1991)}]{Bowen1991}
{Bowen}, G.~H. \& {Willson}, L.~A. 1991, \apjl, 375, L53

\bibitem[{{Boyer} {et~al.}(2012){Boyer}, {Srinivasan}, {Riebel}, {McDonald},
  {van Loon}, {Clayton}, {Gordon}, {Meixner}, {Sargent}, \&
  {Sloan}}]{Boyer2012}
{Boyer}, M.~L., {Srinivasan}, S., {Riebel}, D., {et~al.} 2012, \apj, 748, 40

\bibitem[{{Buchanan} {et~al.}(2006){Buchanan}, {Kastner}, {Forrest}, {Hrivnak},
  {Sahai}, {Egan}, {Frank}, \& {Barnbaum}}]{Buchanan2006}
{Buchanan}, C.~L., {Kastner}, J.~H., {Forrest}, W.~J., {et~al.} 2006, \aj, 132,
  1890

\bibitem[{{Cami}(2002)}]{Cami2002}
{Cami}, J. 2002, PhD thesis, University of Amsterdam

\bibitem[{{de Vries} {et~al.}(2010){de Vries}, {Min}, {Waters}, {Blommaert}, \&
  {Kemper}}]{deVries2010}
{de Vries}, B.~L., {Min}, M., {Waters}, L.~B.~F.~M., {Blommaert}, J.~A.~D.~L.,
  \& {Kemper}, F. 2010, \aap, 516, A86

\bibitem[{{Demyk} {et~al.}(2000){Demyk}, {Dartois}, {Wiesemeyer}, {Jones}, \&
  {d'Hendecourt}}]{Demyk2000}
{Demyk}, K., {Dartois}, E., {Wiesemeyer}, H., {Jones}, A.~P., \&
  {d'Hendecourt}, L. 2000, \aap, 364, 170

\bibitem[{{Dijkstra} {et~al.}(2005){Dijkstra}, {Speck}, {Reid}, \&
  {Abraham}}]{Dijkstra2005}
{Dijkstra}, C., {Speck}, A.~K., {Reid}, R.~B., \& {Abraham}, P. 2005, \apjl,
  633, L133

\bibitem[{{Dorschner} {et~al.}(1995){Dorschner}, {Begemann}, {Henning},
  {Jaeger}, \& {Mutschke}}]{Dorschner1995}
{Dorschner}, J., {Begemann}, B., {Henning}, T., {Jaeger}, C., \& {Mutschke}, H.
  1995, \aap, 300, 503

\bibitem[{{Draine} \& {Lee}(1984)}]{Draine1984}
{Draine}, B.~T. \& {Lee}, H.~M. 1984, \apj, 285, 89

\bibitem[{{Egan} {et~al.}(2001){Egan}, {Van Dyk}, \& {Price}}]{Egan2001}
{Egan}, M.~P., {Van Dyk}, S.~D., \& {Price}, S.~D. 2001, \aj, 122, 1844

\bibitem[{{Feautrier}(1964)}]{Feautrier1964}
{Feautrier}, P. 1964, Comptes Rendus Academie des Sciences (serie non
  specifiee), 258, 3189

\bibitem[{{Fluks} {et~al.}(1994){Fluks}, {Plez}, {The}, {de Winter},
  {Westerlund}, \& {Steenman}}]{Fluks1994}
{Fluks}, M.~A., {Plez}, B., {The}, P.~S., {et~al.} 1994, \aaps, 105, 311

\bibitem[{{Gail} \& {Sedlmayr}(1999)}]{GailSed1999}
{Gail}, H. \& {Sedlmayr}, E. 1999, \aap, 347, 594

\bibitem[{{Gail}(2010)}]{Gail2010}
{Gail}, H.-P. 2010, in Lecture Notes in Physics, Berlin Springer Verlag, Vol.
  815, Lecture Notes in Physics, Berlin Springer Verlag, ed. T.~{Henning},
  61--141

\bibitem[{{Gielen} {et~al.}(2011){Gielen}, {Bouwman}, {van Winckel}, {Lloyd
  Evans}, {Woods}, {Kemper}, {Marengo}, {Meixner}, {Sloan}, \&
  {Tielens}}]{Gielen2011}
{Gielen}, C., {Bouwman}, J., {van Winckel}, H., {et~al.} 2011, \aap, 533, A99

\bibitem[{{Groenewegen}(2006)}]{Groenewegen2006}
{Groenewegen}, M.~A.~T. 2006, \aap, 448, 181

\bibitem[{{Groenewegen} {et~al.}(2009){Groenewegen}, {Sloan}, {Soszy{\'n}ski},
  \& {Petersen}}]{Groenewegen2009}
{Groenewegen}, M.~A.~T., {Sloan}, G.~C., {Soszy{\'n}ski}, I., \& {Petersen},
  E.~A. 2009, \aap, 506, 1277

\bibitem[{{Habing} \& {Olofsson}(2003)}]{HabingBook}
{Habing}, H.~J. \& {Olofsson}, H., eds. 2003, {Asymptotic giant branch stars}
  ({Springer})

\bibitem[{{Hackwell}(1972)}]{Hackwell1972}
{Hackwell}, J.~A. 1972, \aap, 21, 239

\bibitem[{{Heras} \& {Hony}(2005)}]{Heras2005}
{Heras}, A.~M. \& {Hony}, S. 2005, \aap, 439, 171

\bibitem[{{H{\"o}fner}(2007)}]{Hoefner2007}
{H{\"o}fner}, S. 2007, in Astronomical Society of the Pacific Conference
  Series, Vol. 378, Why Galaxies Care About AGB Stars: Their Importance as
  Actors and Probes, ed. F.~{Kerschbaum}, C.~{Charbonnel}, \& R.~F. {Wing}, 145

\bibitem[{{H{\"o}fner}(2008)}]{Hoefner2008}
{H{\"o}fner}, S. 2008, \aap, 491, L1

\bibitem[{{Hron} {et~al.}(1997){Hron}, {Aringer}, \& {Kerschbaum}}]{Hron1997}
{Hron}, J., {Aringer}, B., \& {Kerschbaum}, F. 1997, \aap, 322, 280

\bibitem[{{Ita} {et~al.}(2008){Ita}, {Onaka}, {Kato}, {Tanab{\'e}}, {Sakon},
  {Kaneda}, {Kawamura}, {Shimonishi}, {Wada}, {Usui}, {Koo}, {Matsuura},
  {Takahashi}, {Nakada}, {Hasegawa}, \& {Tamura}}]{Ita2008}
{Ita}, Y., {Onaka}, T., {Kato}, D., {et~al.} 2008, \pasj, 60, 435

\bibitem[{{Ivezic} \& {Elitzur}(1995)}]{Ivezic1995}
{Ivezic}, Z. \& {Elitzur}, M. 1995, \apj, 445, 415

\bibitem[{{Ivezic} \& {Elitzur}(1997)}]{Ivezic1997}
{Ivezic}, Z. \& {Elitzur}, M. 1997, \mnras, 287, 799

\bibitem[{{J{\"a}ger} {et~al.}(2003){J{\"a}ger}, {Fabian}, {Schrempel},
  {Dorschner}, {Henning}, \& {Wesch}}]{Jager2003}
{J{\"a}ger}, C., {Fabian}, D., {Schrempel}, F., {et~al.} 2003, \aap, 401, 57

\bibitem[{{Jones} {et~al.}(2012){Jones}, {Kemper}, {Sargent}, {McDonald},
  {Gielen}, {Woods}, {Sloan}, {Boyer}, {Zijlstra}, {Clayton}, {Kraemer},
  {Srinivasan}, \& {Ruffle}}]{Jones2012}
{Jones}, O.~C., {Kemper}, F., {Sargent}, B.~A., {et~al.} 2012, \mnras, 427,
  3209

\bibitem[{{Kastner} {et~al.}(2008){Kastner}, {Thorndike}, {Romanczyk},
  {Buchanan}, {Hrivnak}, {Sahai}, \& {Egan}}]{Kastner2008}
{Kastner}, J.~H., {Thorndike}, S.~L., {Romanczyk}, P.~A., {et~al.} 2008, \aj,
  136, 1221

\bibitem[{{Kato} {et~al.}(2012){Kato}, {Ita}, {Onaka}, {Tanab{\'e}},
  {Shimonishi}, {Sakon}, {Kaneda}, {Kawamura}, {Wada}, {Usui}, {Koo},
  {Matsuura}, \& {Takahashi}}]{Kato2012}
{Kato}, D., {Ita}, Y., {Onaka}, T., {et~al.} 2012, \aj, 144, 179

\bibitem[{{Kemper} {et~al.}(2002){Kemper}, {de Koter}, {Waters}, {Bouwman}, \&
  {Tielens}}]{Kemper2002}
{Kemper}, F., {de Koter}, A., {Waters}, L.~B.~F.~M., {Bouwman}, J., \&
  {Tielens}, A.~G.~G.~M. 2002, \aap, 384, 585

\bibitem[{{Kemper} {et~al.}(2001){Kemper}, {Waters}, {de Koter}, \&
  {Tielens}}]{Kemper2001}
{Kemper}, F., {Waters}, L.~B.~F.~M., {de Koter}, A., \& {Tielens}, A.~G.~G.~M.
  2001, \aap, 369, 132

\bibitem[{{Kemper} {et~al.}(2010){Kemper}, {Woods}, {Antoniou}, {Bernard},
  {Blum}, {Boyer}, {Chan}, {Chen}, {Cohen}, {Dijkstra}, {Engelbracht},
  {Galametz}, {Galliano}, {Gielen}, {Gordon}, {Gorjian}, {Harris}, {Hony},
  {Hora}, {Indebetouw}, {Jones}, {Kawamura}, {Lagadec}, {Lawton}, {Leisenring},
  {Madden}, {Marengo}, {Matsuura}, {McDonald}, {McGuire}, {Meixner}, {Mulia},
  {O'Halloran}, {Oliveira}, {Paladini}, {Paradis}, {Reach}, {Rubin},
  {Sandstrom}, {Sargent}, {Sewilo}, {Shiao}, {Sloan}, {Speck}, {Srinivasan},
  {Szczerba}, {Tielens}, {van Aarle}, {Van Dyk}, {van Loon}, {Van Winckel},
  {Vijh}, {Volk}, {Whitney}, {Wilkins}, \& {Zijlstra}}]{Kemper2010}
{Kemper}, F., {Woods}, P.~M., {Antoniou}, V., {et~al.} 2010, \pasp, 122, 683

\bibitem[{{Koike} {et~al.}(1995){Koike}, {Kaito}, {Yamamoto}, {Shibai},
  {Kimura}, \& {Suto}}]{Koike1995}
{Koike}, C., {Kaito}, C., {Yamamoto}, T., {et~al.} 1995, \icarus, 114, 203

\bibitem[{{Ku{\v c}inskas} {et~al.}(2006){Ku{\v c}inskas}, {Hauschildt},
  {Brott}, {Vansevi{\v c}ius}, {Lindegren}, {Tanab{\'e}}, \&
  {Allard}}]{Kucinskas2006}
{Ku{\v c}inskas}, A., {Hauschildt}, P.~H., {Brott}, I., {et~al.} 2006, \aap,
  452, 1021

\bibitem[{{Ku{\v c}inskas} {et~al.}(2005){Ku{\v c}inskas}, {Hauschildt},
  {Ludwig}, {Brott}, {Vansevi{\v c}ius}, {Lindegren}, {Tanab{\'e}}, \&
  {Allard}}]{Kucinskas2005}
{Ku{\v c}inskas}, A., {Hauschildt}, P.~H., {Ludwig}, H.-G., {et~al.} 2005,
  \aap, 442, 281

\bibitem[{{Lebzelter} {et~al.}(2006){Lebzelter}, {Posch}, {Hinkle}, {Wood}, \&
  {Bouwman}}]{Lebzelter2006}
{Lebzelter}, T., {Posch}, T., {Hinkle}, K., {Wood}, P.~R., \& {Bouwman}, J.
  2006, \apjl, 653, L145

\bibitem[{{Little-Marenin} \& {Little}(1990)}]{Little-Marenin1990}
{Little-Marenin}, I.~R. \& {Little}, S.~J. 1990, \aj, 99, 1173

\bibitem[{{Marshall} {et~al.}(2004){Marshall}, {van Loon}, {Matsuura}, {Wood},
  {Zijlstra}, \& {Whitelock}}]{Marshall2004}
{Marshall}, J.~R., {van Loon}, J.~T., {Matsuura}, M., {et~al.} 2004, \mnras,
  355, 1348

\bibitem[{{Mathis} {et~al.}(1977){Mathis}, {Rumpl}, \&
  {Nordsieck}}]{Mathis1977}
{Mathis}, J.~S., {Rumpl}, W., \& {Nordsieck}, K.~H. 1977, \apj, 217, 425

\bibitem[{{McDonald} {et~al.}(2010){McDonald}, {Sloan}, {Zijlstra},
  {Matsunaga}, {Matsuura}, {Kraemer}, {Bernard-Salas}, \&
  {Markwick}}]{McDonald2010}
{McDonald}, I., {Sloan}, G.~C., {Zijlstra}, A.~A., {et~al.} 2010, \apjl, 717,
  L92

\bibitem[{{Min} {et~al.}(2003){Min}, {Hovenier}, \& {de Koter}}]{Min2003}
{Min}, M., {Hovenier}, J.~W., \& {de Koter}, A. 2003, \aap, 404, 35

\bibitem[{{Monnier} {et~al.}(1998){Monnier}, {Geballe}, \&
  {Danchi}}]{Monnier1998}
{Monnier}, J.~D., {Geballe}, T.~R., \& {Danchi}, W.~C. 1998, \apj, 502, 833

\bibitem[{{Mutschke} {et~al.}(1998){Mutschke}, {Begemann}, {Dorschner},
  {Guertler}, {Gustafson}, {Henning}, \& {Stognienko}}]{Mutschke1998}
{Mutschke}, H., {Begemann}, B., {Dorschner}, J., {et~al.} 1998, \aap, 333, 188

\bibitem[{{Norris} {et~al.}(2012){Norris}, {Tuthill}, {Ireland}, {Lacour},
  {Zijlstra}, {Lykou}, {Evans}, {Stewart}, \& {Bedding}}]{Norris2012}
{Norris}, B.~R.~M., {Tuthill}, P.~G., {Ireland}, M.~J., {et~al.} 2012, \nat,
  484, 220

\bibitem[{{Olofsson}(2004)}]{Olofsson2004}
{Olofsson}, H. 2004, in {Asymptotic Giant Branch Stars}, ed. H.~J. {Habing} \&
  H.~{Olofsson}, 325

\bibitem[{{Onaka} {et~al.}(1989){Onaka}, {de Jong}, \& {Willems}}]{Onaka1989}
{Onaka}, T., {de Jong}, T., \& {Willems}, F.~J. 1989, \aap, 218, 169

\bibitem[{{Ordal} {et~al.}(1988){Ordal}, {Bell}, {Alexander}, {Newquist}, \&
  {Querry}}]{Ordal1988}
{Ordal}, M.~A., {Bell}, R.~J., {Alexander}, Jr., R.~W., {Newquist}, L.~A., \&
  {Querry}, M.~R. 1988, \ao, 27, 1203

\bibitem[{{Ossenkopf} {et~al.}(1992){Ossenkopf}, {Henning}, \&
  {Mathis}}]{Ossenkopf1992}
{Ossenkopf}, V., {Henning}, T., \& {Mathis}, J.~S. 1992, \aap, 261, 567

\bibitem[{{Pietrzy{\'n}ski} {et~al.}(2013){Pietrzy{\'n}ski}, {Graczyk},
  {Gieren}, {Thompson}, {Pilecki}, {Udalski}, {Soszy{\'n}ski}, {Koz{\l}owski},
  {Konorski}, {Suchomska}, {Bono}, {Moroni}, {Villanova}, {Nardetto},
  {Bresolin}, {Kudritzki}, {Storm}, {Gallenne}, {Smolec}, {Minniti}, {Kubiak},
  {Szyma{\'n}ski}, {Poleski}, {Wyrzykowski}, {Ulaczyk}, {Pietrukowicz},
  {G{\'o}rski}, \& {Karczmarek}}]{Pietrzynski2013}
{Pietrzy{\'n}ski}, G., {Graczyk}, D., {Gieren}, W., {et~al.} 2013, \nat, 495,
  76

\bibitem[{{Posch} {et~al.}(2002){Posch}, {Kerschbaum}, {Mutschke}, {Dorschner},
  \& {J{\"a}ger}}]{Posch2002}
{Posch}, T., {Kerschbaum}, F., {Mutschke}, H., {Dorschner}, J., \& {J{\"a}ger},
  C. 2002, \aap, 393, L7

\bibitem[{{Riebel} {et~al.}(2012){Riebel}, {Srinivasan}, {Sargent}, \&
  {Meixner}}]{Riebel2012}
{Riebel}, D., {Srinivasan}, S., {Sargent}, B., \& {Meixner}, M. 2012, \apj,
  753, 71

\bibitem[{{Sargent} {et~al.}(2011){Sargent}, {Srinivasan}, \&
  {Meixner}}]{Sargent2011}
{Sargent}, B.~A., {Srinivasan}, S., \& {Meixner}, M. 2011, \apj, 728, 93

\bibitem[{{Sargent} {et~al.}(2010){Sargent}, {Srinivasan}, {Meixner}, {Kemper},
  {Tielens}, {Speck}, {Matsuura}, {Bernard}, {Hony}, {Gordon}, {Indebetouw},
  {Marengo}, {Sloan}, \& {Woods}}]{Sargent2010}
{Sargent}, B.~A., {Srinivasan}, S., {Meixner}, M., {et~al.} 2010, \apj, 716,
  878

\bibitem[{{Sloan} {et~al.}(2003){Sloan}, {Kraemer}, {Goebel}, \&
  {Price}}]{Sloan2003}
{Sloan}, G.~C., {Kraemer}, K.~E., {Goebel}, J.~H., \& {Price}, S.~D. 2003,
  \apj, 594, 483

\bibitem[{{Sloan} {et~al.}(2008){Sloan}, {Kraemer}, {Wood}, {Zijlstra},
  {Bernard-Salas}, {Devost}, \& {Houck}}]{Sloan2008}
{Sloan}, G.~C., {Kraemer}, K.~E., {Wood}, P.~R., {et~al.} 2008, \apj, 686, 1056

\bibitem[{{Sloan} {et~al.}(2010){Sloan}, {Matsunaga}, {Matsuura}, {Zijlstra},
  {Kraemer}, {Wood}, {Nieusma}, {Bernard-Salas}, {Devost}, \&
  {Houck}}]{Sloan2010}
{Sloan}, G.~C., {Matsunaga}, N., {Matsuura}, M., {et~al.} 2010, \apj, 719, 1274

\bibitem[{{Sloan} \& {Price}(1995)}]{Sloan1995}
{Sloan}, G.~C. \& {Price}, S.~D. 1995, \apj, 451, 758

\bibitem[{{Sloan} \& {Price}(1998)}]{Sloan1998}
{Sloan}, G.~C. \& {Price}, S.~D. 1998, \apjs, 119, 141

\bibitem[{{Sogawa} \& {Kozasa}(1999)}]{Sogawa1999}
{Sogawa}, H. \& {Kozasa}, T. 1999, \apjl, 516, L33

\bibitem[{{Speck} {et~al.}(2000){Speck}, {Barlow}, {Sylvester}, \&
  {Hofmeister}}]{Speck2000}
{Speck}, A.~K., {Barlow}, M.~J., {Sylvester}, R.~J., \& {Hofmeister}, A.~M.
  2000, \aaps, 146, 437

\bibitem[{{Srinivasan} {et~al.}(2011){Srinivasan}, {Sargent}, \&
  {Meixner}}]{Srinivasan2011}
{Srinivasan}, S., {Sargent}, B.~A., \& {Meixner}, M. 2011, \aap, 532, A54

\bibitem[{{Stencel} {et~al.}(1990){Stencel}, {Nuth}, {Little-Marenin}, \&
  {Little}}]{Stencel1990}
{Stencel}, R.~E., {Nuth}, III, J.~A., {Little-Marenin}, I.~R., \& {Little},
  S.~J. 1990, \apjl, 350, L45

\bibitem[{{Suh}(1999)}]{Suh1999}
{Suh}, K.-W. 1999, \mnras, 304, 389

\bibitem[{{Tielens}(1990)}]{Tielens1990}
{Tielens}, A.~G.~G.~M. 1990, in From Miras to Planetary Nebulae: Which Path for
  Stellar Evolution?, ed. {M.~O.~Mennessier \& A.~Omont} ({Editions
  Frontieres}), 186--200

\bibitem[{{Treffers} \& {Cohen}(1974)}]{TreffersCohen1974}
{Treffers}, R. \& {Cohen}, M. 1974, \apj, 188, 545

\bibitem[{{Ueta} \& {Meixner}(2003)}]{Ueta2003}
{Ueta}, T. \& {Meixner}, M. 2003, \apj, 586, 1338

\bibitem[{{van Loon} {et~al.}(2001){van Loon}, {Zijlstra}, {Bujarrabal}, \&
  {Nyman}}]{vanLoon2001}
{van Loon}, J.~T., {Zijlstra}, A.~A., {Bujarrabal}, V., \& {Nyman}, L.-{\AA}.
  2001, \aap, 368, 950

\bibitem[{{Vassiliadis} \& {Wood}(1993)}]{Vassiliadis1993}
{Vassiliadis}, E. \& {Wood}, P.~R. 1993, \apj, 413, 641

\bibitem[{{Verhoelst} {et~al.}(2009){Verhoelst}, {van der Zypen}, {Hony},
  {Decin}, {Cami}, \& {Eriksson}}]{Verhoelst2009}
{Verhoelst}, T., {van der Zypen}, N., {Hony}, S., {et~al.} 2009, \aap, 498, 127

\bibitem[{{Whitelock} {et~al.}(1994){Whitelock}, {Menzies}, {Feast}, {Marang},
  {Carter}, {Roberts}, {Catchpole}, \& {Chapman}}]{Whitelock1994}
{Whitelock}, P., {Menzies}, J., {Feast}, M., {et~al.} 1994, \mnras, 267, 711

\bibitem[{{Whitelock} {et~al.}(2003){Whitelock}, {Feast}, {van Loon}, \&
  {Zijlstra}}]{Whitelock2003}
{Whitelock}, P.~A., {Feast}, M.~W., {van Loon}, J.~T., \& {Zijlstra}, A.~A.
  2003, \mnras, 342, 86

\bibitem[{{Wood} {et~al.}(1992){Wood}, {Whiteoak}, {Hughes}, {Bessell},
  {Gardner}, \& {Hyland}}]{Wood1992}
{Wood}, P.~R., {Whiteoak}, J.~B., {Hughes}, S.~M.~G., {et~al.} 1992, \apj, 397,
  552

\bibitem[{{Woods} {et~al.}(2011){Woods}, {Oliveira}, {Kemper}, {van Loon},
  {Sargent}, {Matsuura}, {Szczerba}, {Volk}, {Zijlstra}, {Sloan}, {Lagadec},
  {McDonald}, {Jones}, {Gorjian}, {Kraemer}, {Gielen}, {Meixner}, {Blum},
  {Sewi{\l}o}, {Riebel}, {Shiao}, {Chen}, {Boyer}, {Indebetouw}, {Antoniou},
  {Bernard}, {Cohen}, {Dijkstra}, {Galametz}, {Galliano}, {Gordon}, {Harris},
  {Hony}, {Hora}, {Kawamura}, {Lawton}, {Leisenring}, {Madden}, {Marengo},
  {McGuire}, {Mulia}, {O'Halloran}, {Olsen}, {Paladini}, {Paradis}, {Reach},
  {Rubin}, {Sandstrom}, {Soszy{\'n}ski}, {Speck}, {Srinivasan}, {Tielens}, {van
  Aarle}, {van Dyk}, {van Winckel}, {Vijh}, {Whitney}, \&
  {Wilkins}}]{Woods2010}
{Woods}, P.~M., {Oliveira}, J.~M., {Kemper}, F., {et~al.} 2011, \mnras, 411,
  1597

\bibitem[{{Woolf} \& {Ney}(1969)}]{WoolfNey1969}
{Woolf}, N.~J. \& {Ney}, E.~P. 1969, \apjl, 155, L181

\bibitem[{{Wright} {et~al.}(2010){Wright}, {Eisenhardt}, {Mainzer}, {Ressler},
  {Cutri}, {Jarrett}, {Kirkpatrick}, {Padgett}, {McMillan}, {Skrutskie},
  {Stanford}, {Cohen}, {Walker}, {Mather}, {Leisawitz}, {Gautier}, {McLean},
  {Benford}, {Lonsdale}, {Blain}, {Mendez}, {Irace}, {Duval}, {Liu}, {Royer},
  {Heinrichsen}, {Howard}, {Shannon}, {Kendall}, {Walsh}, {Larsen}, {Cardon},
  {Schick}, {Schwalm}, {Abid}, {Fabinsky}, {Naes}, \& {Tsai}}]{Wright2010}
{Wright}, E.~L., {Eisenhardt}, P.~R.~M., {Mainzer}, A.~K., {et~al.} 2010, \aj,
  140, 1868

\end{thebibliography}

 \end{document}